\documentstyle[preprint,prd,tighten,eqsecnum,aps,epsf]{revtex}
\begin{document}
\preprint{\vbox{\hbox{SUNY-NTG-94-16}
\hbox{May 1994}
\hbox{Phys. Rev. D {\bf 51}, 6017 (1995)}
\hbox{      }\hbox{      }\hbox{       }
}}
\draft
\title{\bf Formalism for dilepton production
via virtual photon bremsstrahlung in hadronic reactions}
\author{Peter Lichard\footnote{
On leave of absence from the 
Department of Theoretical Physics, Faculty of Mathematics and Physics,
Comenius University, SK 842-15 Bratislava, Slovakia.
}}
\address{Department of Physics, State University of New York 
at Stony Brook,\\
Stony Brook, New York 11794}
 
\maketitle
\begin{abstract}
We derive a set of new formulas for various distributions in dilepton 
production via virtual photon bremsstrahlung from pseudoscalar mesons
and unpolarized spin-one-half fermions. These formulas correspond to 
the leading and sub-leading terms in the Low--Burnett--Kroll expansion 
for real photon bremsstrahlung. The relation of our leading-term 
formulas to previous works is also shown. Existing formulas are 
examined in the light of Lorentz covariance and gauge invariance. 
Numerical comparison is made in a simple example, where an ``exact" 
formula and real photon data exist. The results reveal large 
discrepancies among different bremsstrahlung formulas. Of all the 
leading-term bremsstrahlung formulas, the one derived in this work 
agrees best with the exact formula. The issues of $M_T$-scaling and 
event generators are also addressed.
\end{abstract}

 
\narrowtext

\section{INTRODUCTION}

Even if electromagnetic phenomena rank among the best understood in
particle physics, their merging with hadronic processes often brings
ambiguities that cannot be resolved on a purely theoretical
basis. It is generally believed, however, that many of the reactions 
in which photons and dileptons are produced can be described as 
bremsstrahlung from incoming and outgoing charged particles, at least
within a limited kinematic range.
 
It is well known that the cross section for production of photons 
with very low energies is uniquely determined by the cross section of
the corresponding nonradiative reaction 
\cite{low,adlerdot,burkroll,brown,bellroy,duca}. In the dilepton 
sector, the situation is less clear. Several different bremsstrahlung 
formulas have been proposed 
\cite{rueckl,craithom,goshawee,galkap,galkap89,bpp,clgore1,hagaem2}, 
which, as will be demonstrated in this work, do not agree with one 
another very well. Some of these formulas fail to satisfy constraints 
implied by general principles such as Lorentz covariance and gauge 
invariance. Besides the formulas cited above, a few additional 
formulas exist that have been designed for specific kinematic regions 
\cite{farrfrau,craigie}. They will not be considered here. 

The purpose of this work is to present a set of consistent formulas 
for various distributions in dilepton bremsstrahlung from pseudoscalar
mesons and unpolarized spin-one-half fermions. We consider the terms 
that are proportional to the square of the nonradiative matrix element 
(leading term approximation) and its derivatives (sub-leading 
approximation). We first derive the formula for the most general 
quantity, namely the double differential cross section in the momenta
of leptons. The correct form for it has not been yet known even in the
leading term approximation. Then we arrive at the cross section in 
dilepton mass and momentum. Its leading part differs only by 
higher-order terms in the dilepton four-momentum from one of the 
already existing formulas \cite{bpp}.

To stress the importance of using the correct virtual bremsstrahlung
formalism, we compare our formulas to those already existing in the 
literature. The comparison is made first on general grounds and is 
followed by an application of the formulas to a simple physical 
process. All formulas are scrutinized from the point of view of 
Lorentz covariance \cite{einstein} and from what we will call a global 
variable test.

The global variable test is based on the finding \cite{baier,invint} 
that in a one-photon approximation, gauge invariance leads to the 
following relation for the inclusive\footnote{We will omit the word 
inclusive in what follows. All our cross sections for photon or 
dilepton production are semi-inclusive (integrated over the final 
state hadron momenta in a reaction with given number and types
of final state particles). Some of the relations are more general 
and hold also for inclusive cross sections, which are given as sums 
of the semi-inclusive cross sections over all possible reactions with 
chosen initial particles [as, e.g., (\ref{global}), which is valid
even for exclusive cross sections]. In some cases, the sum over all 
possible reactions should be supplemented to make the relation 
inclusive.} differential cross section in global dilepton quantities 
$M$ (dilepton mass) and ${\bf q}$ (dilepton momentum)
\begin{equation}
\label{global}
\frac{d^4\sigma^{e^+e^-}}{dM^2d^3q} =
{\cal T}(M^2)\ 
\left(\frac{d^3\sigma^{\gamma^*}}{d^3q}\right)_M
\end{equation}
with the function ${\cal T}$ given by
\begin{equation}
\label{tmsq}
{\cal T}(M^2) =
\frac{\alpha}{3\pi}\frac{1}{M^2}
\left(1+\frac{2\mu^2}{M^2}\right)
\sqrt{1-\frac{4\mu^2}{M^2}}\ ,
\end{equation}
where $\alpha$ is the fine structure constant and $\mu$ the lepton 
mass. The rightmost quantity in Eq.~(\ref{global}) is called the 
cross section for virtual photon
production. It does not have a direct physical meaning, as it is not
experimentally accessible. Nevertheless, Eqs.~(\ref{global}) and
(\ref{tmsq}) can test the soundness of theoretical formulas, 
because they show the only two places in which the lepton mass $\mu$ 
may and must (unless neglected) appear. From a technical point of 
view, the virtual photon cross section is calculable more easily than 
the dilepton cross section \cite{baier}. Of course, if one needs the 
distribution in the momenta of leptons, a more involved approach is 
unavoidable. 
 
Furthermore, if an otherwise identical reaction exists in which a 
photon is produced instead of a dilepton (this need not always be the 
case, viz., $\pi^+\pi^-\rightarrow e^+e^-$), the relation
\begin{equation}
\label{limit}
\lim_{M\rightarrow 0}
\left(\frac{d^3\sigma^{\gamma^*}}{d^3q}\right)_M
=\frac{d^3\sigma^{\gamma}}{d^3q}
\end{equation}
must be fulfilled \cite{baier} with the differential cross section 
for real photon production on the right-hand side. 

Relations analogous to (\ref{global}) and (\ref{limit}) can also
be written for the corresponding quantities in other than two-body 
reactions, namely for decays and processes with more than two 
particles in the initial state \cite{lichard}.

In the next section, we derive the formulas appropriate for the 
production of very soft (low mass, low momentum) dileptons via 
virtual photon bremsstrahlung in reactions with charged pseudoscalar 
particles. Section~\ref{fermions} deals with the virtual 
bremsstrahlung from fermions. 
In Sec.~\ref{survey}, we show the additional approximations which 
lead to the various formulas that have appeared explicitly or as 
a part of more complex expressions in the literature. All the 
formulas are examined to see if they fulfill the Lorentz covariance
and global variable tests. In Sec.~\ref{model}, we introduce a simple 
theoretical model of the process $\rho^0\rightarrow \pi^+\pi^-e^+e^-$
(which has not been experimentally investigated yet) based on 
a successful description \cite{singer} of the recently observed 
\cite{vasserman} decay $\rho^0\rightarrow \pi^+\pi^-\gamma$. The 
former process will be a testing ground for various virtual 
bremsstrahlung formulas in Sec.~\ref{bremsstrahlung} with the 
theoretical distribution from Sec.~\ref{model} serving as 
a reference. We summarize our main points and add a few comments in 
Sec.~\ref{comments}. Some related issues
are discussed in the Appendices. In Appendix \ref{appmt} we
show how $M_T$-scaling transpires from the leading term virtual
bremsstrahlung formalism. Appendix \ref{appsingvirt} deals with
the ``exact" formula for the $\rho^0\rightarrow \pi^+\pi^-
\gamma^*$ branching ratio. 
Appendix \ref{appleading} 
addresses the issue of photon and dilepton event generators that 
conserve energy and momentum. 

\section{Dileptons from virtual 
bremsstrahlung off pseudoscalar mesons}
\label{mesons}

In this section, we assume that all charged particles are pseudoscalar
mesons. The technically more involved case of virtual bremsstrahlung 
from spin-one-half fermions will be treated in the next section.

\subsection{Leading term approximation}
\label{leading}
\begin{figure}
\begin{center}
\leavevmode
\setlength \epsfxsize{16.5cm}
\epsffile{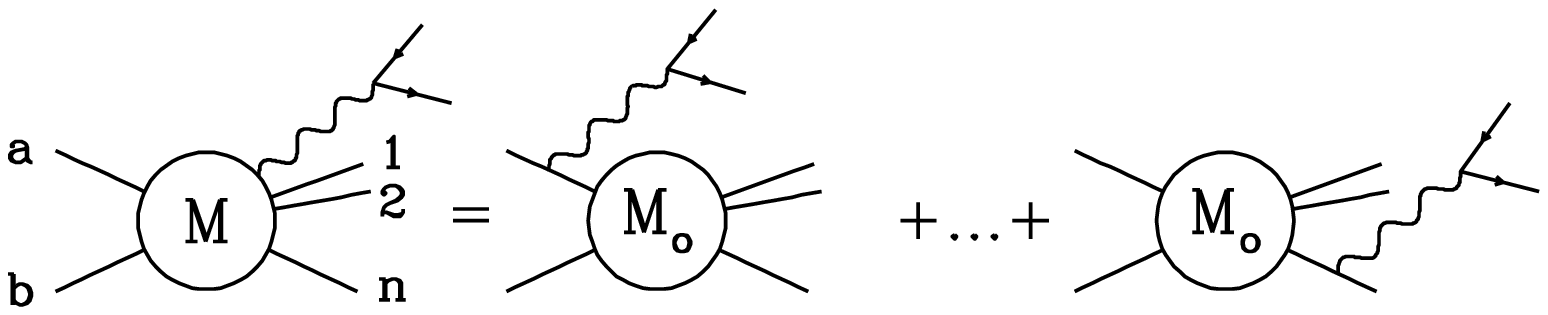}
\end{center}
\caption
{Matrix element for dilepton production in virtual bremsstrahlung
approximation.}
\label{me}
\end{figure}

Let us consider a $2\rightarrow n$ hadronic reaction 
\begin{equation}
\label{reaction1}
a+b\rightarrow 1+2+\cdots+n
\end{equation} 
and denote its matrix element as 
${\cal M}_0\equiv {\cal M}_0(p_a,p_b,p_1,\cdots,p_{n})$. Our aim is to
find the matrix element ${\cal M}$ of the reaction 
\begin{equation}
\label{reaction2}
a+b\rightarrow 1+2+\cdots+n+l^++l^-,
\end{equation} 
in which a soft lepton pair with the four-momentum $q=p_++p_-$
is produced in addition to $n$ hadrons. The dominant contribution
to the matrix element ${\cal M}$ comes from diagrams where the virtual
photon is attached to one of the external legs (see Fig.~\ref{me}).  
The diagrams in which a virtual photon is radiated from internal lines 
give subleading contributions. This is caused by the nonvanishing 
virtuality contributions of the type $(p^2-m^2)$ to the denominators 
of newly emerging propagators.  Such terms do not appear if one of the
two particles attached to the electromagnetic vertex is real 
($p^2=m^2$). See below.

Using the Feynman rules of pseudoscalar electrodynamics
\cite{bjorken} we can immediately write down the contribution to the 
matrix element ${\cal M}$ from radiation of an initial ($x=a,b$)
\begin{equation}
\label{minitial}
{\cal M}_x=-eQ_x{\cal M}_0\ \frac{(2p_x-q)^\mu}
{2\,p_x\!\cdot\!q-M^2}L_\mu
\end{equation}
and a final ($i=1,\cdots,n$) 
\begin{equation}
\label{mfinal}
{\cal M}_i=eQ_i{\cal M}_0\ \frac{(2p_i+q)^\mu}
{2\,p_i\!\cdot\!q+M^2}L_\mu
\end{equation}
state particle. Above, $e$ is the positive elementary charge, 
$Q_x$ ($Q_i$) is the charge of an initial (a final) particle, and $M$
is the dilepton mass ($q^2=M^2$). As is customary in this field, we 
have supposed that the nonradiative matrix element does not change 
when an incoming or outgoing momentum becomes ``slightly" 
off-mass-shell; for example,
\begin{equation}
\label{offshell}
{\cal M}_0(p_a-q,p_b,p_1,\cdots,p_{n})=
{\cal M}_0(p_a,p_b,p_1,\cdots,p_{n})\equiv {\cal M}_0\ .
\end{equation}
The lepton part is given by
\begin{equation}
\label{leptonterm}
L_\mu=\frac{e}{M^2}\bar u^{(s_-)}(p_-)\gamma_\mu v^{(s_+)}(p_+)\ .
\end{equation}
With this choice of $L_\mu$, the matrix element for photon 
production is obtained by the substitution 
$L_\mu\rightarrow\epsilon_\mu$, where $\epsilon$ is the photon 
polarization vector. Summing the contributions of individual diagrams, 
we arrive at
\begin{equation}
\label{msum}
{\cal M}=e{\cal M}_0\ J\!\cdot\!L
\end{equation}
with
\begin{equation}
\label{j}
J^\mu=-\sum_{x=a,b} Q_x\frac{(2p_x-q)^\mu}{2\,p_x\!\cdot\!q-M^2}+
      \sum_{i=1}^{n} Q_i\frac{(2p_i+q)^\mu}{2\,p_i\!\cdot\!q+M^2}\ .
\end{equation}
The four-vector $J$ satisfies the important relation
\begin{equation}
\label{jq}
J\!\cdot\!q=-Q_a-Q_b+\sum_{i=1}^{n} Q_i=0\ ,
\end{equation}
which reflects charge conservation. Squaring the matrix 
element~(\ref{msum}) and summing over the spins of leptons, we get
\begin{equation}
\label{summsq1}
\sum_{s_+,s_-}\left|{\cal M}\right|^2 = 4\pi\alpha\left|
{\cal M}_0\right|^2\ J^\mu J^\nu\ L_{\mu\nu}\ .
\end{equation}
The tensor $L_{\mu\nu}$ is defined by
\begin{equation}
\label{lmunudef}
L_{\mu\nu}=\sum_{s_+,s_-}\ L_\mu L_\nu^*\ .
\end{equation}
A straightforward calculation leads to
\begin{equation}
\label{lmunu}
L_{\mu\nu}=\frac{8\pi\alpha}{M^4}
\left(q_\mu q_\nu-l_\mu l_\nu-M^2 g_{\mu\nu}\right)\ ,
\end{equation}
where we have introduced the four-vector $l=p_+-p_-$ as the difference 
of leptons' momenta. Using the above relation we can 
rewrite~(\ref{summsq1}) in the form
\begin{equation}
\label{summsq2}
\sum_{s_+,s_-}\left|{\cal M}\right|^2 = \left|{\cal M}_0\right|^2
\frac{32\pi\alpha^2}{M^2}\left[-J^2-\frac{1}{M^2}
\left(l\!\cdot\!J\right)^2
\right].
\end{equation}
Inserting this into the relation for the unpolarized cross section 
of the reaction (\ref{reaction2}) leads to 
\begin{eqnarray}
\label{xsec}
d\sigma & = &\frac{1}{4E_aE_b|{\bf v}_a-{\bf v}_b|}
\left|{\cal M}_0\right|^2
(2\pi)^4\delta(p_a+p_b-\sum_i p_i-q)
\nonumber \\[5pt]
& & \times
\frac{32\pi\alpha^2}{M^2}\left[-J^2-\frac{1}{M^2}
\left(l\!\cdot\!J\right)^2
\right]\ 
\prod_{i=1}^n\frac{d^3p_i}{2E_i(2\pi)^3}
\prod_{l=+,-}\frac{d^3p_l}{2E_l(2\pi)^3}\ .
\end{eqnarray}
After neglecting the dilepton four-momentum in the argument of the 
$\delta$-function and integrating over the momenta of final hadrons, 
we get
\begin{equation}
\label{d6sig}
E_+E_-\frac{d^6\sigma^{e^+e^-}}{d^3p_+d^3p_-}=\frac{\alpha^2}
{8\pi^4}\frac{1}
{M^2}\int\left[-J^2-\frac{1}{M^2}(l\!\cdot\!J)^2\right]d\sigma_0\ ,
\end{equation}
where
\begin{equation}
\label{xsec0}
d\sigma_0 = \frac{1}{4E_aE_b|{\bf v}_a-{\bf v}_b|}
\left|{\cal M}_0\right|^2
(2\pi)^4\delta(p_a+p_b-\sum_i p_i)
\prod_{i=1}^n\frac{d^3p_i}{2E_i(2\pi)^3}
\end{equation}
is the infinitesimal cross section of the reaction (\ref{reaction1}).
Double differential cross section 
$E_+E_-d^6\sigma^{e^+e^-}/d^3p_+d^3p_-$
is the most general quantity that characterizes the production of pairs
of unpolarized unlike-sign leptons. Knowing it, we can find any other
distribution (but not {\it vice versa}).

A little exercise from relativistic kinematics provides us with
the following general formula
\begin{equation}
\label{page45}
E\ \frac{d^6\sigma^{e^+e^-}}{dM^2d^3q\ d\tilde\Omega_+} =
\frac{1}{4}\sqrt{1-\frac{4\mu^2}{M^2}}\,\, 
E_+E_-\frac{d^6\sigma^{e^+e^-}}{d^3p_+d^3p_-}\ ,
\end{equation}
where $E=E_++E_-$ is the dilepton energy and $d\tilde\Omega_+$ is the
solid angle element for positron momentum in the dilepton rest frame. 
Using (\ref{d6sig}) and (\ref{page45}), we arrive at another formula 
of our virtual bremsstrahlung formalism
\begin{equation}
\label{d6sigmix}
E\ \frac{d^6\sigma^{e^+e^-}}{dM^2d^3q\ d\tilde\Omega_+} =
\frac{\alpha^2}{32\pi^4}\frac{1}{M^2}
\sqrt{1-\frac{4\mu^2}{M^2}}
\int\left[-J^2-\frac{1}{M^2}(l\!\cdot\!J)^2\right]d\sigma_0\ .
\end{equation}
Finally, integrating over the positron angles in the dilepton rest 
frame, we obtain the differential cross section in global dilepton 
variables
\begin{equation}
\label{newglobal}
E\frac{d^4\sigma^{e^+e^-}}{dM^2d^3q} =
\frac{\alpha^2}{12\pi^3}\frac{1}{M^2}
\left(1+\frac{2\mu^2}{M^2}\right)
\sqrt{1-\frac{4\mu^2}{M^2}}
\int\left(-J^2\right)\ d\sigma_0\ .
\end{equation}
This is just formula (\ref{global}) with the virtual photon cross 
section given by
\begin{equation}
\label{myvirtual}
E\ \left(\frac{d^3\sigma^{\gamma^*}}{d^3q}\right)_M
=\frac{\alpha}{4\pi^2}\int\left(-J^2\right)\ d\sigma_0\ .
\end{equation}
Sometimes it is advantageous to express $J^2$ in terms of 
three-dimensional
vectors. This can easily be done \cite{bpp} using the relation
$J_0=E^{-1}\ {\bf J}\!\cdot\!{\bf q}$,
which follows from Eq.~(\ref{jq}). We thus get
\begin{equation}
\label{j2}
-J^2=\left({\bf J}\times{\bf n}\right)^2+\frac{M^2}{E^2}
\left({\bf J}\!\cdot\!{\bf n}\right)^2,
\end{equation}
where ${\bf n} = {\bf q}/|{\bf q}|$ is the unit vector in the dilepton
momentum direction.

To investigate the limit (\ref{limit}) of (\ref{myvirtual}), 
it is sufficient to realize that due to (\ref{j2}), the part of 
$J^\mu$ that is proportional to $q^\mu$ will not contribute in the 
limit $M\rightarrow 0$. What remains can be written as

\begin{equation}
\label{myreal} 
\omega\ \frac{d^3\sigma^\gamma}{d^3q}
=\frac{\alpha}{4\pi^2}\int\left(-J_R^2\right)\ d\sigma_0 
\end{equation}
with $\omega=|{\bf q}|$ and
\begin{equation}
\label{jr}
J_R = \sum  Q^\prime_i\frac{p_i}{p_i\!\cdot\!q}
        =\frac{1}{\omega}\sum  \frac{Q^\prime_i}{E_i}\ \frac{p_i}
{1-{\bf v}_i\!\cdot\!{\bf n}}\ .
\end{equation}
To make the formula more compact, we have introduced the variable
$ Q^\prime_i$, which is identical with the charge of
final particles and acquires the opposite sign for initial particles.
The unspecified sum runs over all (initial and final) hadrons
and ${\bf v}_i={\bf p}_i/E_i$ is the velocity of the $i$th hadron.
Eqs.~(\ref{myreal}) and (\ref{jr}) combine to give the well known
leading term formula for real photon bremsstrahlung 
\cite{jackson,berest}.

The central results of this subsection are the formulas (\ref{d6sig}),
(\ref{d6sigmix}), and (\ref{newglobal}) for various differential 
cross sections of dilepton production via virtual photon 
bremsstrahlung in the leading term approximation.

\subsection{Next-to-leading term approximation}
\label{nexttoleading}
When going beyond the next-to-leading order, the nonradiative matrix 
element is no longer considered to be immune against the changes in 
incoming and outgoing momenta. We write, instead of 
Eq.~(\ref{offshell}),
\begin{equation}
\label{m0xq}
{\cal M}_0(p_x-q)={\cal M}_0-
q^\alpha\frac{\partial{\cal M}_0}{\partial p_x^\alpha}
\end{equation}
if one of the incoming momenta changes, and
\begin{equation}
\label{m0iq}
{\cal M}_0(p_i+q)={\cal M}_0+
q^\alpha\frac{\partial{\cal M}_0}{\partial p_i^\alpha}
\end{equation}
to account for a change in one of the outgoing momenta. In the above 
two equations, we suppress the momenta that keep their ``nonradiative"
values. If we now simply incorporated the ``corrected" values 
(\ref{m0xq}) and (\ref{m0iq}) into the expressions (\ref{minitial}) 
and (\ref{mfinal}) for the radiative matrix elements, and summed these
up, we would obtain a non-gauge-invariant quantity. After replacing 
the four-vector $L_\mu$ by the virtual photon four-momentum,
it would not vanish. The reason is that we have not yet included the 
``contact" terms, which are generated from the strong interaction 
Lagrangian by the minimal electromagnetic interaction principle.
The latter says that the electromagnetic interaction terms appear as 
a result of the substitution
\begin{equation}
\label{minelpr}
p^\alpha\rightarrow p^\alpha-eQg^{\alpha\mu}\ ,
\end{equation}
where the index $\mu$ is to be contracted with the real photon
polarization vector or the virtual photon propagator. To
find the contact terms in our case, we make a formal expansion of
${\cal M}_0(p^\alpha_i-eQ_ig^{\alpha\mu})$, where $p_i$ denotes any
of the incoming or outgoing momenta. We thus find that the contact 
term associated with the $i$th hadron is
\begin{equation}
\label{contacti}
C_i^\mu=-eQ_i\frac{\partial{\cal M}_0}{\partial p_{i\mu}}\ .
\end{equation}
For the same reason as stated in the previous subsection, the 
radiation from internal lines will not contribute in this 
approximation either. Putting it all together, the radiative matrix 
element comes out to be
\begin{equation}
\label{mek}
{\cal M}=e\ K\!\cdot\!L\ ,
\end{equation}
where $L$ is the four-vector defined by Eq.~(\ref{leptonterm}), and
\begin{equation}
\label{kmu}
K^\mu={\cal M}_0\ J^\mu+\sum_{i}Q_i\frac{\partial {\cal M}_0}
{\partial p_i^\alpha}\left(\frac{p_i^\mu q^\alpha}{p_i\!\cdot\!q}-
g^{\mu\alpha}\right).
\end{equation}
We retained only the leading and next-to-leading terms in $q$. The 
sum runs over both incoming and outgoing hadrons and the four-vector 
$J$ is given by Eq.~(\ref{j}). To get the cross section for producing
a pair of unpolarized leptons, we need the quantity 
\begin{equation}
\label{summsqh}
\sum_{s_+,s_-}\left|{\cal M}\right|^2 = 4\pi\alpha
\  H^{\mu\nu} L_{\mu\nu}\ ,
\end{equation}
with symmetric tensor $L_{\mu\nu}$ defined by (\ref{lmunudef}), and
\begin{equation}
\label{hdef}
H^{\mu\nu}=\frac{1}{2}\left(K^\mu {K^*}^\nu+{K^*}^\mu K^\nu\right)\ .
\end{equation} 
Keeping only the $q$-terms of the same order as before, we easily 
obtain
\begin{eqnarray}
\label{hmunu}
H^{\mu\nu}&=&\left|{\cal M}_0\right|^2\ J^\mu J^\nu +\frac{1}{2}
\sum_{i,j}\frac{Q_i Q_j^\prime}{(p_i\!\cdot\!q)(p_j\!\cdot\!q)}
\frac{\partial\left|{\cal M}_0\right|^2}{\partial p_i^\beta}
p_{i\alpha}\nonumber \\
& &\times
\left[p_j^\mu\left(g^{\nu\alpha}q^\beta-g^{\nu\beta}q^\alpha\right)+
p_j^\nu\left(g^{\mu\alpha}q^\beta-g^{\mu\beta}q^\alpha\right)
\right]\ .
\end{eqnarray}
To simplify the formula, we have again used the convention 
$Q_j^\prime=-Q_j$ for
the incoming hadrons and $Q_j^\prime=Q_j$ for the outgoing hadrons.
The differential cross section in lepton pair momenta now reads
\begin{eqnarray}
\label{d6sigsub}
E_+E_-\frac{d^6\sigma^{e^+e^-}}{d^3p_+d^3p_-}&=&\frac{\alpha^2}
{8\pi^4}\frac{1}{M^2}\biggl\{
\int\Bigl[-J^2-\frac{1}{M^2}(l\!\cdot\!J)^2\Bigr]\ d\sigma_0\ + 
\  \sum_{i,j}\int\frac{Q_iQ_j^\prime}{M^2(p_i\!\cdot\!q)
(p_j\!\cdot\!q)}\biggr.\nonumber \\
& & \times \biggl.
\Bigl[p_i\!\cdot\!q\left(l\!\cdot\!p_j\ l^\beta+M^2 p_j^\beta\right)
-\left(l\!\cdot\!p_i\ l\!\cdot\!p_j+M^2 p_i\!\cdot\!p_j\right)q^\beta
\Bigr]\ d\sigma_{0,i\beta}\biggr\}.
\end{eqnarray}
The notation is same as we met in Eq.~(\ref{d6sig}), except for
\begin{equation}
\label{xsec0ibeta}
d\sigma_{0,i\beta} = \frac{1}{4E_aE_b|{\bf v}_a-{\bf v}_b|}
\frac{\partial\left|{\cal M}_0\right|^2}{\partial p^\beta_i}
(2\pi)^4\delta(p_a+p_b-\sum_k p_k)
\prod_{k=1}^n\frac{d^3p_k}{2E_k(2\pi)^3}\ .
\end{equation}

Using the same procedure as in subsection \ref{leading}, we arrive at 
the differential cross section in global dilepton variables in the 
form~(\ref{global}) with the virtual photon cross section given by
\begin{eqnarray}
\label{virtsubl}
E\ \left(\frac{d^3\sigma^{\gamma^*}}{d^3q}\right)_M
&=&\frac{\alpha}{4\pi^2}\biggr\{\int\left(-J^2\right)\ d\sigma_0\ + 
\  \sum_{i,j}\int\frac{Q_iQ_j^\prime}{(p_i\!\cdot\!q)(p_j\!\cdot\!q)}
\biggr.\nonumber \\
& & \times \biggl.
\Bigl[(p_i\!\cdot\!q) p_j^\beta
-(p_i\!\cdot\!p_j)q^\beta\Bigr]\ d\sigma_{0,i\beta}\biggr\}.
\end{eqnarray}
We will comment on the zero photon mass limit of this equation
in Sec.~\ref{fermions}.

\subsection{Decays}
\label{decays}
The description of virtual bremsstrahlung in hadronic decays can be 
achieved following the lines sketched for the two-body reactions. 
We need only
to change the number of incoming particles to one and replace the
cross sections by decay widths. This follows from the similar 
structure of the relations between cross section, or decay width, on 
the one hand and the matrix element squared on the other. Another 
important factor is the universality (with respect to the numbers of 
incoming and outgoing particles) of Eqs.~(\ref{msum}) and (\ref{j}). 

Let us consider the decay
\begin{equation}
\label{decay}
a\rightarrow 1+2+\cdots+n+l^++l^-.
\end{equation}
For its differential decay width we can write, in the leading term
approximation,
\begin{equation}
\label{d6decay}
E_+E_-\frac{d^6\Gamma^{e^+e^-}}{d^3p_+d^3p_-}=\frac{\alpha^2}
{8\pi^4}\frac{1}
{M^2}\int\left[-J^2-\frac{1}{M^2}(l\!\cdot\!J)^2\right]d\Gamma_0\ .
\end{equation}
The quantity
\begin{equation}
\label{gamma0}
d\Gamma_0 = \frac{1}{2m_a}\left|{\cal M}_0\right|^2
(2\pi)^4\delta(p_a-\sum_i p_i)
\prod_{i=1}^n\frac{d^3p_i}{2E_i(2\pi)^3}
\end{equation}
is the invariant decay width into an infinitesimal element of the 
momentum space for the decay
\begin{equation}
\label{decay0}
a\rightarrow 1+2+\cdots+n\ .
\end{equation}
The other leading-term formulas can be easily modified as
well. The differential decay width in global dilepton quantities is
\begin{equation}
\label{globalwidth}
E\frac{d^4\Gamma^{e^+e^-}}{dM^2d^3q} =
{\cal T}(M^2)\ 
\frac{\alpha}{4\pi^2}\int\left(-J^2\right)\ d\Gamma_0\ .
\end{equation}
If a nonradiative decay (\ref{decay0}) contains only two particles
in the final state, we can proceed further to get
\begin{equation}
\label{restwidth}
\frac{1}{\Gamma_0}\frac{d^2\Gamma^{e^+e^-}}{dM^2dq^*} =
{\cal T}(M^2)\ \frac{\alpha}{4\pi^2}
\frac{{q^*}^2}{E^*}\int\left(-J^2\right)\ d\Omega_{{\bf q}^{*}}\ ,
\end{equation}
where the asterisk refers to quantities in the rest frame of the 
parent particle and $q^* = |{\bf q}^{\,*}|$. For a hypothetical decay 
into two particles and a virtual photon it means
\begin{equation}
\label{virtdecay}
\frac{1}{\Gamma_0}\left(\frac{d\Gamma^{\gamma^*}}{dq^*}\right)_M =
\frac{\alpha}{4\pi^2}
\frac{{q^*}^2}{E^*}\int\left(-J^2\right)\ d\Omega_{{\bf q}^{*}}\ .
\end{equation}

\section{Virtual bremsstrahlung from fermions}
\label{fermions}

It has been shown in \cite{bpp} that for virtual bremsstrahlung
from fermions, the terms proportional
to $p^\mu$ in the expression (\ref{j}) can be obtained in the same 
wayas in the real photon bremsstrahlung off fermions (see, e.g., 
\cite{berest}). In order to obtain the form of $J$ analogous to
(\ref{j}), additional approximations are required. One has to discard
some terms while keeping others of the same order in $q$. The
justification of such a procedure is unclear.

One may hope that if the summing (or averaging) over spins of hadrons 
is performed, the additional contributions will rearrange and the 
formulas identical to those pertaining to the pseudoscalar case will 
be restored. This would resemble a similar development in the 
next-to-leading terms in real photon bremsstrahlung \cite{burkroll}. 

To investigate such a possibility, let as assume that in the 
nonradiative reaction (\ref{reaction1}), a charged fermion-antifermion 
(e.g., proton-antiproton) pair is produced. To be more definite, we 
assign the index 1 to the fermion ($Q_1=1$) and 2 to the antifermion 
($Q_2=-1$). Now, the matrix element exhibits the form
\begin{equation}
\label{mef1}
{\cal M}_0=\bar u^{(s_1)}(p_1)\ \Gamma\ v^{(s_2)}(p_2)\ ,
\end{equation}
where $\Gamma\equiv\Gamma(p_a,p_b,p_1,\cdots,p_n)$ is a matrix in the
spinor space. To simplify notation, we will display its arguments 
only if they differ from the values just shown.
Squaring the matrix element (\ref{mef1}) and summing over the spin 
projections $s_1$ and $s_2$, we get
\begin{equation}
\label{mef1sq}
\overline{\left|{\cal M}_0\right|^2}={\rm Tr}\left[(\hat{p}_1+m)\Gamma
(\hat{p}_2-m)\Gamma^\prime\right]\ ,
\end{equation}
where $\Gamma^\prime=\gamma^0\Gamma^\dagger\gamma^0$. The quantity 
(\ref{mef1sq}) determines the unpolarized cross section of the 
nonradiative reaction (\ref{reaction1}) with the two spin-one-half 
hadrons in the final state. For a later use, let us notice that
\begin{equation}
\label{dm0sqdp1}
\frac{\partial{\overline{\left|{\cal M}_0\right|^2}}}
{\partial p_1^\alpha}={\rm Tr}\left\{\left[
\Gamma^\prime\gamma_\alpha\Gamma+
\Gamma^\prime(\hat{p}_1+m)\frac{\partial\Gamma}{\partial p_1^\alpha}+
\frac{\partial\Gamma^\prime}{\partial p_1^\alpha}
(\hat{p}_1+m)\Gamma\right](\hat{p}_2-m)\right\}\ ,
\end{equation}
and
\begin{equation}
\label{dm0sqdp2}
\frac{\partial{\overline{\left|{\cal M}_0\right|^2}}}
{\partial p_2^\alpha}={\rm Tr}\left\{
(\hat{p}_1+m)\left[\Gamma\gamma_\alpha\Gamma^\prime+
\frac{\partial\Gamma}{\partial p_2^\alpha}
(\hat{p}_2-m)\Gamma^\prime+
\Gamma(\hat{p}_2-m)
\frac{\partial\Gamma^\prime}{\partial p_2^\alpha}
\right]\right\}\ .
\end{equation}

As a next step, let us consider the corresponding dilepton-producing 
reaction (\ref{reaction2}). We will concentrate on virtual 
bremsstrahlung from fermions and, for simplicity, take all the mesons 
neutral. In addition, we neglect any anomalous electromagnetic
interactions. The changes in the strong interaction matrix element 
will be incorporated by 
\begin{equation}
\label{gamma12q}
\Gamma(p_i+q)=\Gamma+q^\alpha\frac{\partial\Gamma}
{\partial p_i^\alpha}\ .
\end{equation}
The contact electromagnetic interaction term associated with $i$th 
fermion line leaving the strong interaction core comes out as
\begin{equation}
\label{contfer}
C_i^\mu=-eQ_i\frac{\partial\Gamma}{\partial p_{i\mu}}\ .
\end{equation}
Using the Feynman rules for spinor electrodynamics, 
Dirac equation in momentum space, and the properties of the 
$\gamma$-matrices, we find the matrix element of reaction 
(\ref{reaction2})
\begin{equation}
\label{mef2}
{\cal M} = eL_\mu\ \bar u^{(s_1)}(p_1)\ K^\mu v^{(s_2)}(p_2)\ .
\end{equation}
The four-vector $L_\mu$ has the same meaning as before 
[see (\ref{leptonterm})].
We have, neglecting higher than linear $q$-terms in the numerators,
\begin{equation}
\label{kfermion}
K^\mu=\Gamma J^\mu+
\frac{1}{4\ p_1\!\cdot\!q}\biglb[\gamma^\mu,\hat{q}\bigrb]\Gamma+
\frac{1}{4\ p_2\!\cdot\!q}\Gamma\biglb[\gamma^\mu,\hat{q}\bigrb]+
\left(g^{\mu\alpha}q^\beta-g^{\mu\beta}q^\alpha\right)
\sum_{i=1,2}Q_i\frac{p_{i,\alpha}}{p_i\!\cdot\!q}
\frac{\partial\Gamma}{\partial p_i^\beta} \ .
\end{equation}
In accordance with (\ref{j}), we denoted
\begin{equation}
\label{j12}
J^\mu=\frac{(2p_1+q)^\mu}{2\,p_1\!\cdot\!q+M^2}-
      \frac{(2p_2+q)^\mu}{2\,p_2\!\cdot\!q+M^2}\ .
\end{equation}
At this point we can clearly see the difference between pseudoscalar
and fermion cases. When we neglected the changes in the strong 
interaction core as well as the contact terms in pseudoscalar case, 
we obtained immediately the leading term approximation in the form 
(\ref{msum}). For fermions, the extra terms 
(those with commutators) prevent us from reaching the same goal.

The sum over the spins of the matrix element (\ref{mef2}) squared 
assumes 
the form
\begin{equation}
\label{mef2sq}
\sum_{s_1,s_2,s_+,s_-}\left|{\cal M}\right|^2=4\pi\alpha\ H^{\mu\nu}
L_{\mu\nu}\ .
\end{equation}
The tensor $L_{\mu\nu}$ is given by Eq.~(\ref{lmunu}). Up to the 
leading and next-to-leading order in $q$, 
\begin{eqnarray}
\label{hmunuftr}
H^{\mu\nu}&=&\overline{\left|{\cal M}_0\right|^2}\ J^\mu J^\nu+
\frac{1}{2}\Bigl\{
J^\mu\ {\rm Tr}\bigl[\left(\hat{p}_1+m\right)\Gamma
\left(\hat{p}_2-m\right){K^\prime}^\nu\bigr]\Bigr.\nonumber\\
& &+\Bigl. J^\mu\ {\rm Tr}\bigl[\left(\hat{p}_1+m\right)K^\nu
\left(\hat{p}_2-m\right)\Gamma^\prime\bigr]+\ \mu \leftrightarrow \nu
\Bigr\}\ 
\end{eqnarray}
with ${K^\prime}^\nu=\gamma_0(K^\nu)^\dagger\gamma_0$.
A straightforward manipulation guides us to an expression in parts of 
which we are able to identify the right-hand sides of 
Eqs.~(\ref{dm0sqdp1}) and (\ref{dm0sqdp2}). After replacing them by 
corresponding derivatives we get
\begin{eqnarray}
\label{hmunuf}
H^{\mu\nu}&=&\overline{\left|{\cal M}_0\right|^2}\ J^\mu J^\nu +
\frac{1}{2}
\sum_{i,j}\frac{Q_i Q_j^\prime}{(p_i\!\cdot\!q)(p_j\!\cdot\!q)}
p_{i\alpha}\frac{\partial\overline{\left|{\cal M}_0\right|^2}}
{\partial p_i^\beta}\nonumber \\
& &\times\left[p_j^\mu\left(g^{\nu\alpha}q^\beta-g^{\nu\beta}q^\alpha
\right)+p_j^\nu\left(g^{\mu\alpha}q^\beta-g^{\mu\beta}q^\alpha\right)
\right]\ .
\end{eqnarray}
This is identical with what we would get from Eq.~(\ref{hmunu}) for 
two outgoing charged mesons, with only one difference. A simple 
square of nonradiative mesonic matrix element is replaced by the sum 
over fermion spins. As an independent check of our result, we explored 
also other situations (an incoming fermion-antifermion pair, one 
incoming--one outgoing fermion/antifermion) and reached the same 
conclusion. For the  initial state fermions, the sum is replaced by 
the average. The generalization to more than one fermion pair is 
obvious.

The tensor $H^{\mu\nu}$ is the central object of all bremsstrahlung
formulas. We have therefore proven that the dilepton production via 
virtual bremsstrahlung off unpolarized spin-one-half fermions is 
governed, in the leading and next-to-leading approximation, by the 
same formulas as that off pseudoscalar mesons.

Especially, if we neglect the terms proportional to derivatives of 
the unpolarized nonradiative matrix element squared, the tensor 
$H^{\mu\nu}$ reduces to
\begin{equation}
\label{hmunufac}
H^{\mu\nu}=J^\mu J^\nu\ \overline{\left|{\cal M}_0\right|^2}
\end{equation}
with $J^\mu$ given by Eq.~(\ref{j}).
The key leading term approximation relations (\ref{d6sig}), 
(\ref{d6sigmix}), and (\ref{newglobal}), which are of most practical
interest, are valid also for virtual bremsstrahlung from unpolarized 
fermions.

It is a good check that our expression (\ref{virtsubl}) for the
virtual photon cross section meets, in the limit of zero photon mass,
the unpolarized photon cross section calculated from the Burnett and 
Kroll~\cite{burkroll} matrix element.

\section{A survey of virtual bremsstrahlung formulas}
\label{survey}
\subsection{R\"{u}ckl formula}
In his work \cite{rueckl}, R\"{u}ckl suggested the
formula
\begin{equation}
\label{rueckl}
E_+E_-\frac{d^6\sigma^{e^+e^-}}{d^3p_+d^3p_-}=\frac{\alpha}{2\pi^2}
\frac{1}{M^2}\left(\omega\frac{d^3\sigma^\gamma}{d^3q}
\right)_{{\bf q}={\bf p}_++{\bf p}_-},
\end{equation}
which links the cross section for production of dileptons via virtual 
photon 
bremsstrahlung to the bremsstrahlung cross section for real photons.
The meaning of the symbols is as follows: ${\bf p}_+$ and ${\bf p}_-$ 
are the momenta of leptons, $E_\pm=({\bf p}_\pm^{\ 2}+\mu^2)^{1/2}$ 
are their energies, $M^2=(p_++p_-)^2$ is the dilepton mass squared, 
${\bf q}$\,\, is the momentum of photon, $\omega=|{\bf q}|$ is its 
energy, 
and $\alpha$ is the fine-structure constant. Because of the vanishing 
photon mass, it is impossible to satisfy the relation 
$\omega=E_++E_-$ simultaneously with ${\bf q}={\bf p}_++{\bf p}_-$.

Let us investigate now how Eq.~(\ref{rueckl}) copes with general 
principles we mentioned in the Introduction. The principle of 
relativity \cite{einstein} requires that any meaningful formula must 
be relativistically covariant. This implies that if we view the 
bremsstrahlung process from another (primed) frame, we must find the 
same relation among the transformed quantities as we did in the 
original frame: 
\begin{equation}
\label{ruecklprime}
E^\prime_+E^\prime_-\frac{d^6\sigma^{e^+e^-}}{d^3p^\prime_+d^3
p^\prime_-}=\frac{\alpha}{2\pi^2}\frac{1}{M^2}
\left(\omega^\prime\frac{d^3\sigma^\gamma}{d^3q^\prime}
\right)_{{\bf q}^\prime={\bf p}^\prime_++{\bf p}^\prime_-}.
\end{equation}
We have assumed for simplicity that the
velocities of colliding particles are collinear 
and performed a boost along the collision axis (otherwise 
$\sigma$'s also acquire primes). The left-hand sides of 
Eqs.~(\ref{rueckl}) and (\ref{ruecklprime}) are obviously equal. 
Relativistic covariance will thus be satisfied if and only if 
\begin{equation}
\label{covar}
\left(\omega\frac{d^3\sigma^\gamma}{d^3q}
\right)_{{\bf q}={\bf p}_++{\bf p}_-}=
\left(\omega^\prime\frac{d^3\sigma^\gamma}{d^3q^\prime}
\right)_{{\bf q}^\prime={\bf p}^\prime_++{\bf p}^\prime_-}.
\end{equation}
But this condition cannot be fulfilled because in the new frame the 
photon momentum differs from the sum of lepton's momenta. In fact, 
for the longitudinal components of the corresponding vectors we have
\begin{eqnarray}
\label{longtrafo}
p^\prime_{\pm,L}&=&\gamma\left(p_{\pm,L}-\beta E_\pm\right) 
\nonumber\\ q^\prime_L&=&\gamma\left(q_L-\beta \omega\right),
\end{eqnarray} 
which leads to 
\begin{equation}
\label{longmome}
q^\prime_L=p^\prime_{+,L}+p^\prime_{-,L}+\beta\gamma
\left(E_++E_--\omega\right)
\end{equation}
with a nonvanishing extra term on the right-hand side.

In order to apply the global variable test, let us first use 
Eq.~(\ref{page45}) to cast the R\"{u}ckl formula in the form
\begin{equation}
\label{mixrueckl}
E\ \frac{d^6\sigma^{e^+e^-}}{dM^2d^3q\ d\Omega_+^*} =
\frac{\alpha}{8\pi^2}\frac{1}{M^2}\sqrt{1-\frac{4\mu^2}{M^2}}\,\, 
\omega\frac{d^3\sigma^\gamma}{d^3q}\ .
\end{equation}
Integration over the positron momentum angles is simple because
nothing depends on them:
\begin{equation}
\label{globrueckl}
E\ \frac{d^4\sigma^{e^+e^-}}{dM^2d^3q} =
\frac{\alpha}{2\pi}\frac{1}{M^2}\sqrt{1-\frac{4\mu^2}{M^2}}\,\, 
\omega\frac{d^3\sigma^\gamma}{d^3q}\ .
\end{equation}
This formula does not obviously have the form required 
by Eq.~(\ref{global}).
If we nevertheless extract from it the virtual photon cross section, 
which is defined by (\ref{global}), we arrive at
\begin{equation}
\label{virtrueckl}
\left(\frac{d^3\sigma^{\gamma^*}}{d^3q}\right)_M
=\frac{3\omega}{2\sqrt{\omega^2+M^2}}\frac{M^2}{M^2+2\mu^2}
\frac{d^3\sigma^{\gamma}}{d^3q}\ .
\end{equation}
The $M\rightarrow 0$ limit of this expression is zero. If we are not
so strict and require only $\mu\ll M\ll \omega$ (this is the 
situation met, e.g., in the low-mass, high-transverse-momentum 
dielectron production), we are left with another surprising relation
\begin{equation}
\label{1.5rueckl}
\left(\frac{d^3\sigma^{\gamma^*}}{d^3q}\right)_M
=\frac{3}{2}\frac{d^3\sigma^{\gamma}}{d^3q}\ .
\end{equation}
The presence of an incorrect numerical factor in the R\"{u}ckl 
formalism was already noticed by Craigie \cite{craigie}, who
ascribed it to the unjustified omission of the term $l_\mu l_\nu$ in 
the lepton tensor [see Eq.~(\ref{lmunu})].

It is clear that the principal problems we have discussed here are 
not germane to the R\"{u}ckl bremsstrahlung formalism, but are common 
to all the approaches where the dilepton cross section is assumed to 
be proportional  to the photon cross section at the same momentum.

To obtain Eq.~(\ref{rueckl}) in our formalism, we must
(i) write $\ p_i\!\cdot\!q=|{\bf q}\,|E_i-{\bf p}_i\!\cdot\!{\bf q}\ $ 
instead of the correct
$\ p_i\!\cdot\!q=(M^2+{\bf q}^{\ 2})^{1/2}E_i-
{\bf p}_i\!\cdot\!{\bf q}\ $ in Eq.~(\ref{j});
(ii)
omit the term $l_\mu l_\nu$ in Eq.~(\ref{lmunu});
(iii)
neglect $q^\mu$ in the numerators and $M^2$ in the denominators
of Eq.~(\ref{j}).
The first approximation is fatal for Lorentz covariance, the second
one for the global variable test. 

\subsection{Modifications of the R\protect{\"{u}}ckl formula}
\label{modifications}

The R\"{u}ckl formula (\ref{rueckl}) was often used in the 
calculation of dilepton yield in experimental and theoretical works. 
Several modifications of it have been put forward. As will be shown 
later, they brought improvement  in a pragmatic sense, but were not 
able to cure its principal drawbacks.

Gale and Kapusta \cite{galkap} replaced the photon energy squared
$q_0^2$ ($\omega^2$ in our notation) in the denominator in their
Eq.~(5) by a symmetrized combination $E(E^2-M^2)^{1/2}$, where 
$E=E_++E_-$ is the dilepton energy. The quantity $(E^2-M^2)^{1/2}$
represents the dilepton momentum and as such must be equal to
the photon momentum $|{\bf q}|$, which is in turn equal to the
photon energy $q_0$. The replacement $q_0^2\rightarrow 
E(E^2-M^2)^{1/2}$ is thus equivalent to multiplying the right-hand 
side of Eq.~(\ref{rueckl}) by $\omega/E$. The same modification was
used by Haglin, Gale, and Emel'yanov in \cite{hagaem1} and also in 
a part of the paper \cite{clresa91} by Cleymans, Redlich, and Satz.

In a subsequent paper \cite{galkap89}, Gale and Kapusta introduced a 
factor which partially corrected the soft photon approximation for 
processes with two particles in the final state. It accounts for the 
shrinking of the Lorentz invariant phase space available to them, 
which results from the emission of a dilepton with mass $M$ and  
center-of-mass-system energy $E^*$. The
production of dileptons that would violate the energy-momentum
conservation is forbidden. The correction factor is given by
(we display it in a simpler form assuming equal masses $m$ for the
final-state hadrons)
\begin{equation}
\label{rgaka}
R(s,M,E^*)=\sqrt{\frac{s\left(s_2-4m^2\right)}
{s_2\left(s-4m^2\right)}}\ ,
\end{equation}
where $s$ is the invariant energy available for all final-state 
particles and
\begin{equation}
\label{s2}
s_2=s+M^2-2E^*\sqrt{s}\ .
\end{equation}

In a paper \cite{hagaem2} of Haglin, Gale, and
Emel'yanov, the photon energy squared $q_0^2$ enters the 
denominator of the quantity 
$|\epsilon\!\cdot\!J|^2_{ab\rightarrow cd}$
[see their (3.6), (3.7) or (A9)]. It is forced to acquire the value 
of $(E_++E_-)^2$ by the second $\delta$-function in their Eq.~(3.4). 
It induces a multiplicative factor of $\omega^2/E^2$  on the
right-hand side of the R\"{u}ckl formula (\ref{rueckl}). The authors
also used the correction factor (\ref{rgaka}). Recently, Haglin and
Gale \cite{haga94} utilized the same modification of the R\"{u}ckl
formula to assess the bremsstrahlung contribution to the $e^+e^-$
invariant mass distribution in proton-proton and proton-neutron
collisions at the lab kinetic energy of 4.9 GeV. For the final
states with more than two hadrons they modified the phase-space
correction factor accordingly.

\subsection{Paper by Craigie and Thompson}
After a thorough discussion of the real photon bremsstrahlung, the 
authors of~\cite{craithom} turned to dileptons. If they had really 
done what they described verbally at the bottom of p.~129, they would 
have obtained immediately a simple and correct formula for the double 
differential
cross section in the momenta of leptons [our Eq.~(\ref{d6sig})] with
a little different Ansatz for the four-vector $J$ (same as was used 
later in \cite{bpp}). Unfortunately, they instead wrote a cumbersome 
and obviously
wrong formula (3.1) on p.~130. The incorrectness of the latter
can be seen, e.g., from the fact that the quantity Tr$\{\rho L\}$,
which enters it, depends on the momenta of hadrons via the four-vector
$J$ [see their Eq.~(3.2)]. But there is no integration over hadron 
momenta in (3.1) [only that hidden in $2q_0d\sigma/d^3q$, given by 
Eq.~(2.5)]. The left-hand side of formula (3.1) in \cite{craithom} 
thus depends on hadron momenta, which makes it unusable for 
evaluating the inclusive dilepton cross section.

The authors probably tried to express the double differential cross
section as proportional to the cross section for virtual photon 
production. But, as we have  learnt, this is possible only for
the dilepton cross section in global dilepton variables [compare our
(\ref{d6sig}) and (\ref{newglobal})].

\subsection{Formula used by Goshaw {\it et al.}}
The experimentally observed production of very-low-energy
$e^+e^-$ pairs in 18 GeV/$c$ $\pi^\pm p$ collisions was reported
and compared to the expectations based on the leading-term 
bremsstrahlung calculations in the paper \cite{goshawee}. The 
authors used the formula
\begin{eqnarray}
\label{goshaw}
\frac{d\sigma}{d\mu\ d\omega\ d\Omega\ d\Omega^*} &=&\frac{\alpha^2}
{(2\pi)^4}\frac{\left[\mu^2-(2m)^2\right]^{1/2}
\left(\omega^2-\mu^2\right)^{1/2}}{\mu^2\omega^2}\nonumber\\[5pt]
&&\times\int d^3P_3\cdots d^3P_n\left[\frac{(J\!\cdot\!l)^2}
{\mu^2}-(J\!\cdot\!J)
\right]\frac{d\sigma_h}{d^3P_3\ d^3P_4\cdots d^3P_n}\ ,
\end{eqnarray}
where $\mu$, $\omega$, and $d\Omega$ are the dilepton mass, energy, 
and infinitesimal solid angle, respectively, $d\Omega^*$ is the 
infinitesimal solid angle of positron in the dilepton rest frame, 
$l=P_+-P_-$, and
\begin{equation}
\label{jgoshaw1}
J=\omega\sum_{i=1}^n\frac{Q_i}{(P_++P_-)\!\cdot\!P_i}P_i\ .
\end{equation} 
The charge quantum number of outgoing particles ($i=3,\cdots,n$) is 
$Q_i$, of incoming ($i=1,2$) ones $(-Q_i)$. After changing the 
notation used in \cite{goshawee} to ours, Eq.~(\ref{goshaw}) reads
\begin{equation}
\label{d6siggos}
E\ \frac{d^6\sigma^{e^+e^-}}{dM^2d^3q\ d\tilde\Omega_+} =
\frac{\alpha^2}{32\pi^4}\frac{1}{M^2}
\sqrt{1-\frac{4\mu^2}{M^2}}
\int\left[\frac{1}{M^2}(l\!\cdot\!J_G)^2-J_G^2\right]d\sigma_0\ ,
\end{equation}
where now
\begin{equation}
\label{jg}
J_G^\mu=\sum  Q^\prime_i\frac{p_i^\mu}{p_i\!\cdot\!q}\ .
\end{equation}
The formula (\ref{d6siggos}) differs from our Eq.~(\ref{d6sigmix})
by the sign of one of the terms in brackets. It is difficult to trace
the origin of this discrepancy. The derivation of the 
formula (\ref{d6siggos}) has not been published, although it
was signalized in \cite{goshawee}. We suspect that 
Eq.~(\ref{d6siggos}) is not correct, because after integrating it
over the positron momentum directions we get the formula
\begin{equation}
\label{globgosh}
E\ \frac{d^4\sigma^{e^+e^-}}{dM^2d^3q} =
\frac{\alpha^2}{6\pi^3}\frac{1}{M^2}
\left(1-\frac{\mu^2}{M^2}\right)
\sqrt{1-\frac{4\mu^2}{M^2}}\ 
\int (-J_G^2)\ d\sigma_0\ ,
\end{equation}
which does not comply with the global variable test, defined
by Eqs.~(\ref{global}) and (\ref{limit}). In this respect, the fact 
that the authors used additional approximations to get their 
Eq.~(\ref{jgoshaw1}) [compare (\ref{j}) and (\ref{jg})] seems to be 
of lesser importance.

\subsection{Formula of Balek, Pi\v{s}\'{u}tov\'{a}, 
and Pi\v{s}\'{u}t}

In paper \cite{bpp} the formula analogous to our 
(\ref{newglobal}) was written for a charged particle  scattering 
on a neutral particle (or on a potential). Their formalism 
satisfies both the Lorentz covariance and  global variable tests.
However, instead of the vector $J^\mu$ given by Eq.~(\ref{j}), the 
following one was 
chosen:
\begin{equation}
\label{pisutc}
C^\mu=\frac{p_1^\mu}{p_1\!\cdot\!q}-\frac{p_a^\mu}{p_a\!\cdot\!q}\ .
\end{equation}
Here, $p_a$ and $p_1$ are four-momenta of the charged particle 
before and after the scattering, respectively. 
In this case, Eq.~(\ref{j}) becomes
\begin{equation}
\label{jscatt}
J^\mu= \frac{(2p_1+q)^\mu}{2\,p_1\!\cdot\!q+M^2}-
       \frac{(2p_a-q)^\mu}{2\,p_a\!\cdot\!q-M^2}\ .
\end{equation}
To get the Ansatz (\ref{pisutc}) of Balek {\it et~al.} we have to
neglect $q^\mu$ in the numerators and $M^2$ in the denominators above. 

For later convenience we write here the $n$-final-particle
generalization of the Balek {\it et~al.} formula in our notation.
The same conventions as used in (\ref{jr}) apply.
\begin{equation}
\label{bppn}
E\left(\frac{d^3\sigma^{\gamma^*}}{d^3q}\right)_M
=\frac{\alpha}{4\pi^2}\int\ -\left(\sum Q_i^\prime\frac{p_i^\mu}
{E_iE-{\bf p}_i\!\cdot\!{\bf q}\ } \right)^2d\sigma_0\ .
\end{equation}

We will return to the approximation of Balek, Pi\v{s}\'{u}tov\'{a}
and Pi\v{s}\'{u}t in Appendix \ref{appmt}, in connection with the 
concept of transverse-mass scaling (see below).

\subsection{Formula of Cleymans, Goloviznin, and Redlich}
The authors of \cite{clgore1} used the following 
classical-electrodynamics motivated expression for the energy per 
unit of momentum radiated in the form of virtual photons with mass 
$M$ if the charged particle changes its velocity from 
${\bf v}_1$ to ${\bf v}_2$ (we switch from their notation to ours):
\begin{equation}
\label{clgore} 
\frac{d^3I}{d^3q}=\frac{\alpha}{4\pi^2}\left|
\frac{{\bf n} \times{\bf v}_2}{E-{\bf v}_2\!\cdot\!{\bf q}}
-\frac{{\bf n} \times{\bf v}_1}{E-{\bf v}_1\!\cdot\!{\bf q}}
\right|^2.
\end{equation}
Our expression for this quantity stems from the relations
\begin{equation}
\label{enrelation}
\frac{d^3I}{d^3q}=E\frac{d^3N^{\gamma^*}}{d^3q}
               =\frac{1}{\sigma_0}E\left(\frac{d^3\sigma^{\gamma^*}}
                  {d^3q}\right)_M,
\end{equation}
where $N^{\gamma^*}$ is the mean number of virtual photons with mass 
$M$ per a collision, in conjunction with Eqs.~(\ref{j2}) and 
(\ref{myvirtual}). 
Choosing the cross section $d\sigma_0$ that allows only the required 
change of velocity, we get
\begin{equation}
\label{myclgore} 
\frac{d^3I}{d^3q}=\frac{\alpha}{4\pi^2}\left[
\left({\bf n}\times{\bf J}\ \right)^2
+\frac{M^2}{E^2}\left({\bf n}\!\cdot\!{\bf J}\ \right)^2\right]
\end{equation}
with
\begin{equation}
\label{myvectorj}
{\bf J}=\frac{{\bf v}_2+{\bf q}/(2E_2)}
{E-{\bf v}_2\!\cdot\!{\bf q}+M^2/(2E_2)}-
\frac{{\bf v}_1-{\bf q}/(2E_1)}
{E-{\bf v}_1\!\cdot\!{\bf q}-M^2/(2E_1)}.
\end{equation}
Comparing (\ref{clgore}) with (\ref{myclgore}) and (\ref{myvectorj})
we can see that in \cite{clgore1} two additional approximations have
tacitly been made in contrast to our formalism:
(i)
The terms proportional to ${\bf q}\ $ in the numerators and those 
proportional to $M^2$ in the denominators of Eq.~(\ref{myvectorj}) 
have been neglected. This is equivalent to the approximation made 
in~\cite{bpp}.
(ii)
The second term in the brackets of (\ref{myclgore}) has not been 
considered.
This approximation is more dangerous, since it makes the quantity 
(\ref{clgore}) non-covariant under Lorentz transformations. 

\subsection{Transverse mass scaling}

In this and next subsections we are going to report about two 
attempts to relate the virtual photon cross section to the 
experimentally accessible cross section for real photon production.

Farrar and Frautschi \cite{farrfrau} and Cobb {\it et al.} \cite{cobb}
used the concept of transverse mass scaling. They assumed that the 
virtual photon cross section does not depend on three variables 
(longitudinal momentum $q_L$, transverse momentum $q_T$, and virtual 
photon mass $M$), but rather on only two [$q_L$ and transverse mass 
$M_T=(q_T^2+M^2)^{1/2}$]. The condition (\ref{limit}) then leads to
the relation
\begin{equation}
\label{mtscaling}
\left(\frac{d^3\sigma^{\gamma^*}}{d^3q}\right)_M=
\frac{d^3\sigma^\gamma}{d^3p}\ ,
\end{equation}
where $p_L=q_L$ and $p_T=(q_T^2+M^2)^{1/2}$. Because both the energies
and longitudinal momenta of real and virtual photons are now equal,
the relation (\ref{mtscaling}) is covariant under longitudinal Lorentz
boosts. In our test process $\rho^0\rightarrow\pi^+\pi^-e^+e^-$,
this approach will be very successful.

The origin and limitations of the transverse-mass scaling from the 
point of view of the leading term bremsstrahlung formalism are 
discussed in Appendix \ref{appmt}.

\subsection{Real photon approximation}
\label{subsecrpa}

Blockus {\it et al.} \cite{blockus} conjectured that the formula
\begin{equation}
\label{blockus}
\left(E^*\frac{d^3\sigma^{\gamma^*}}{d^3q^*}\right)_M=
\frac{f(M,q_L^*,q_T)}{f(0,q_L^*,q_T)}\ \left(\omega^*
\frac{d^3\sigma^\gamma}{d^3q^*} \right)
\end{equation}
is valid in the center-of-mass frame (we use $q_L^*$ instead of their
$x=2q_L^*/\sqrt{s}$). The quantity $f$ is the structure function for
the production of a virtual photon of mass $M$ summed over photon
polarization states. The authors considered two options for their 
ratio entering the right-hand side of Eq.~(\ref{blockus}): 
(i) independent of $M$ (i.e., identically equal to 1), and 
(ii) linearly dependent on $M$. None of these options seemed to be 
excluded by their data integrated over the region of acceptance. In 
our numerical comparison of various bremsstrahlung formulas we will 
explore the former option 
\begin{equation}
\label{rpa}
\left(E^*\frac{d^3\sigma^{\gamma^*}}{d^3q^*}\right)_M= \omega^*
\frac{d^3\sigma^\gamma}{d^3q^*}\ ,
\end{equation}
and will refer to it as the ``real photon approximation".
 
\section{A model of two-pion radiative decays of $\rho^0$}
\label{model}

As noted earlier, we will check the reliability of the various  
leading term virtual bremsstrahlung formulas by applying them to the 
strong-interaction decay $\rho^0\rightarrow \pi^+\pi^-$ in order to 
get estimates of the differential decay width for
\begin{equation}
\label{rhotoee}
\rho^0\rightarrow \pi^+\pi^-e^+e^-.
\end{equation} 
While a comparison of the bremsstrahlung formulas themselves is 
instructive, additional insight may be gained by also comparing them 
to results from a formalism that goes beyond the leading term 
approximation. To our knowledge, nobody has investigated the 
decay~(\ref{rhotoee}) theoretically yet, probably because the chance 
to detect it experimentally is very meager. We present some estimates 
in Subsection B.

The real photon counterpart of (\ref{rhotoee}), namely 
$\rho^0\rightarrow \pi^+\pi^-\gamma$
has been dealt with by several authors \cite{singer,pascual,renard}. 
In this work we will adopt Singer's approach \cite{singer}, which 
agrees nicely with the experimental data (see below).
We will first recapitulate its main points and then generalize it to 
the massive photon (dilepton) case.
 
To simplify notation, we introduce the following ratio of decay 
widths
\begin{equation}
\label{bgamdef}
B^\gamma=\frac{\Gamma_{\rho^0\rightarrow \pi^+\pi^-\gamma}}
{\Gamma_{\rho^0\rightarrow \pi^+\pi^-}}\ .
\end{equation}
We will call it a branching ratio in spite of a small incorrectness
this introduces ($\Gamma_{\rho^0\rightarrow \pi^+\pi^-}$ is smaller
than the total width of $\rho^0$ by about 1 \%). The quantities
$B^{\gamma^*}$ and $B^{e^+e^-}$ will have analogous meaning. 

\subsection{Decay {$\rho^0\rightarrow \pi^+\pi^-\gamma$}}

Strong interaction dynamics enters Singer's calculation through the 
assumption about the $\rho^0\pi^+\pi^-$ vertex in the form
\begin{equation}
\label{vertex}
V^\alpha=f_{\rho\pi\pi}(p_1-p_2)^\alpha\ ,
\end{equation}
where $p_1$ ($p_2$) is the four-momentum of $\pi^+$ ($\pi^-$). For 
external $\rho$'s, this is to be contracted with the polarization 
vector $\epsilon_\alpha(\lambda_\rho)$. The coupling constant 
$f_{\rho\pi\pi}$ 
can be fixed by utilizing the $\rho\rightarrow\pi\pi$ decay width.

\begin{figure}
\begin{center}
\leavevmode
\setlength \epsfxsize{8cm}
\epsffile{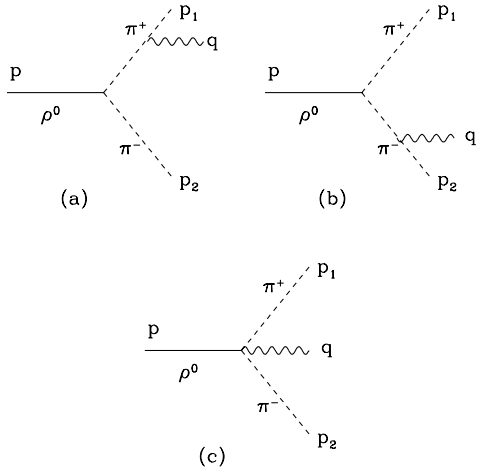}
\end{center}
\caption
{Feynman diagrams for $\rho^0\rightarrow \pi^+\pi^-\gamma$
decay in the approach of Singer \protect\cite{singer}.}
\label{feynman}
\end{figure}

When electromagnetic interactions are switched on by the minimal 
interaction principle, the vertex (\ref{vertex}) generates
a contact term 
\begin{equation}
\label{contact}
C^{\mu\alpha}=-2ef_{\rho\pi\pi}g^{\mu\alpha}\ ,
\end{equation}
which must be considered in addition to the usual vertices
of pseudoscalar electrodynamics (see Fig.~\ref{feynman}). It should
be contracted with both the $\rho$ and $\gamma$ lines which enter the
contact vertex together with the two pion lines. Combining 
the resulting expressions for $\rho^0\rightarrow \pi^+\pi^-$ 
and $\rho^0\rightarrow \pi^+\pi^-\gamma$ decay widths 
from~\cite{singer}, we obtain the formula
\begin{eqnarray}
\label{realsinger}
\frac{dB^\gamma}{d\omega^*}&=&
\frac{4\alpha}{\pi}\frac{1}
{\left(m_\rho^2-4m_\pi^2\right)^{3/2}}\frac{1}{\omega^*}
\biggl\{\omega_m^*\left(m_\rho^2-2m_\pi^2-2m_\rho\omega^*\right)
\ln\frac{1+\xi}{1-\xi}\biggr.\nonumber \\
& & \biggl. -m_\rho\xi
\left[\omega^*_m\left(m_\rho-2\omega^*\right)-2{\omega^*}^2\right]
\biggr\}\ ,
\end{eqnarray}
where $\omega^*$ is the photon energy in the $\rho^0$ rest frame and
\begin{equation}
\label{xi}
\xi=\sqrt\frac{2\left(\omega_m^*-\omega^*\right)}{m_\rho-2\omega^*}.
\end{equation}
The maximum value of $\omega^*$ is 
$\omega_m^*=(m_\rho^2-4m_\pi^2)/(2m_\rho)$. 

\begin{figure}
\begin{center}
\leavevmode
\setlength \epsfxsize{8cm}
\epsffile{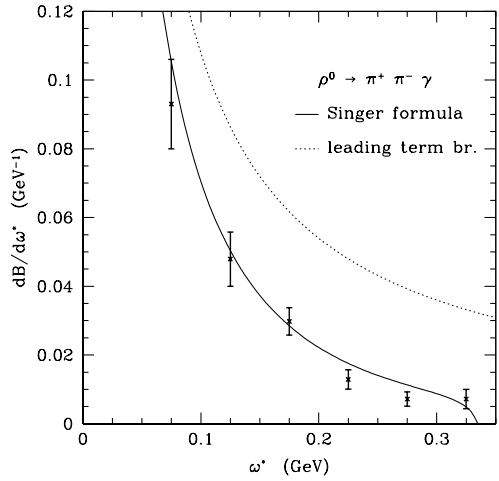}
\end{center}
\caption{The differential branching ratio of 
$\rho^0\rightarrow \pi^+\pi^-\gamma$ as a function of the photon 
energy in the $\rho^0$ rest frame. Solid: Singer formula 
(\protect\ref{realsinger}), dashed: leading
term bremsstrahlung formula (\protect\ref{eq7}). Data 
\protect\cite{vasserman}
were normalized to the integrated branching ratio given in 
\protect\cite{dolinsky}.}
\label{realgam}
\end{figure}

After integrating Eq.~(\ref{realsinger}) over photon energies greater 
than 50 MeV we get the branching ratio of 1.12 \%, in nice agreement 
with the experimental value $(0.99\pm0.16)$ \% \cite{dolinsky}. Also, 
the distribution in photon energies (depicted in Fig.~\ref{realgam} 
by solid curve)
agrees remarkably well with experiment \cite{vasserman}.

\subsection{Decay {$\rho^0\rightarrow \pi^+\pi^-e^+e^-$}}

As has already been stressed, in order to know the differential
decay width in global dilepton variables of the 
decay~(\ref{rhotoee}), it is sufficient to calculate the differential 
decay width for $\rho^0\rightarrow\pi^+\pi^-\gamma^*$ with a massive 
photon. As was proven in \cite{baier}, the evaluation of the latter 
is governed by the same Feynman rules as in the case of a real 
photon. What changes are the kinematical relations which must 
accommodate the non-vanishing photon mass. As a result, we obtain
\begin{eqnarray}
\label{virtsinger}
\left(\frac{dB^{\gamma^*}}{dq^*}\right)_M 
&=&
\frac{4\alpha}{\pi}\frac{1}
{\left(m_\rho^2-4m_\pi^2\right)^{3/2}}
\frac{q^*}{{E^*}^2}\nonumber \\
&\times & \left\{\left[E_m^*\left(m_\rho^2-2m_\pi^2-2E^*m_\rho\right)
+M^2\left(E^*_m+\frac{m_\rho}{4}\right)\right]
\ln\frac{E^*+\xi q^*}{E^*-\xi q^*}\right.\nonumber \\
 &-&\left. E^*q^*\xi
\left[\frac{\left(m_\rho-2E_m^*\right)\left(2m_\rho E^*_m-M^2\right)}
{2\left({E^*}^2-\xi^2{q^*}^2\right)}-2m_\rho\right]\right\}\ .
\end{eqnarray}
Here, 
\begin{equation}
\label{estm}
E_m^*=\frac{m_\rho^2+M^2-4m_\pi^2}{2m_\rho}
\end{equation} 
is the maximum value of $E^*$, the massive photon energy in the 
$\rho^0$-rest frame and
\begin{equation}
\label{ximsq}
\xi=\sqrt\frac{2m_\rho\left(E_m^*-E^*\right)}
{m_\rho^2+M^2-2m_\rho E^*}\ .
\end{equation}
The interested reader may find a few intermediate steps in Appendix
\ref{appsingvirt}. 
Integrating the differential branching ratio for 
$\rho^0\rightarrow\pi^+\pi^-e^+e^-$
\begin{equation}
\label{eesinger}
\frac{d^2B^{e^+e^-}}{dM^2dq^*}={\cal T}(M^2)
\left(\frac{dB^{\gamma^*}}{dq^*}\right)_M
\end{equation}
over the dilepton momenta and masses, we get
the value $1.5\times10^{-4}$ for $\rho^0\rightarrow \pi^+\pi^-e^+e^-$
and $4.9\times10^{-7}$ for $\rho^0\rightarrow \pi^+\pi^-\mu^+\mu^-$.
A low-dielectron-mass cut of 50 MeV/$c^2$ (100 MeV/$c^2$) reduces
the branching ratio to $1.1\times10^{-5}$ ($4.0\times10^{-6}$).

\section{Bremsstrahlung formulas for two-pion
radiative decays of $\rho^0$}
\label{bremsstrahlung}

Our goal here is to calculate the differential branching ratio 
of $\rho^0\rightarrow \pi^+\pi^-e^+e^-$  within all approaches to
dilepton bremsstrahlung we found in the literature. Because some
approaches \cite{bpp,clgore1} do not provide the distribution in the
momenta of electrons, we will concentrate on a less general 
distribution in global dilepton variables. For later use and 
reference we first explore the simpler case of the $\rho^0$ decay 
to a dipion and photon.

\subsection{Decay $\rho^0\rightarrow \pi^+\pi^-\gamma$}
\begin{figure}
\begin{center}
\leavevmode
\setlength \epsfxsize{8cm}
\epsffile{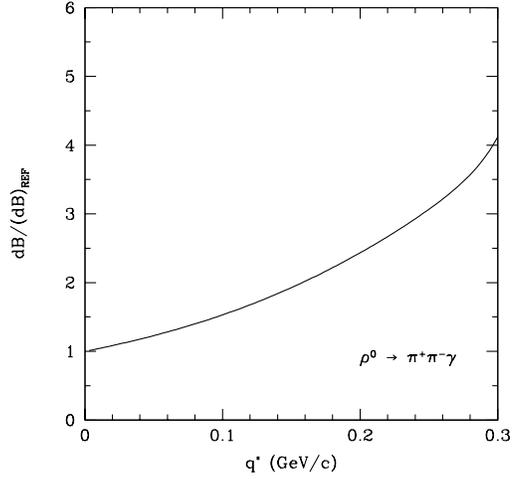}
\end{center}
\caption{The ratio of the differential branching ratio for 
$\rho^0\rightarrow \pi^+\pi^-\gamma$ in leading term approximation
to that of Singer \protect\cite{singer} as a function of the photon 
momentum in the $\rho^0$ rest frame.}
\label{comp000}
\end{figure}

In order to calculate the branching ratio in the leading term 
bremsstrahlung approximation, we start with the formula
\begin{equation}
\label{eq11}
\frac{dB^\gamma}{d\omega^*}=
\frac{\alpha\omega^*}{4\pi^2}\int(-J_R^2)d\Omega_{{\bf q}^*}\ ,
\end{equation}
which is a real photon mutation of Eq.~(\ref{virtdecay}). Using Eqs.
(\ref{j2}) and (\ref{jr}), we can write the integrand in the form
\begin{equation}
\label{jrsqdec}
-J_R^2=\left(\frac{2v^*}{\omega^*}\right)^2\frac{\sin^2\alpha^*}
{\left(1-{v^*}^2\cos^2\alpha^*\right)^2}\ ,
\end{equation}
where
\begin{equation} 
\label{speed}
v^*=\sqrt{1-\frac{4m_\pi^2}{m_\rho^2}}
\end{equation}
is the speed of pions and $\alpha^*$ the angle between the photon and
$\pi^+$ momenta in the $\rho^0$ rest frame. An elementary integration
gives us
\begin{equation}
\label{eq7}
\frac{dB^\gamma}{d\omega^*}=\frac{2\alpha}{\pi\omega^*}
\left(\frac{1+{v^*}^2}{2v^*}
\ln\frac{1+v^*}{1-v^*}-1\right).
\end{equation}
A numerical evaluation shows (Fig.~\ref{realgam}, dashed curve) that 
the leading term bremsstrahlung formula (\ref{eq7}) exceeds the data, 
especially at large momenta. This trend is understandable, as the 
formula ignores the energy-momentum-conservation constraints. The 
branching ratio for $\omega^*>50$ MeV is 2.05 \%, almost twice as 
much as the experimental observation. For later reference, we also 
present in Fig.~\ref{comp000}  the ratio of the 
leading-term-bremsstrahlung
branching ratio (\ref{eq7}) to that calculated from Singer's model
(\ref{realsinger}) as a function of the photon momentum $q^*$ 
(which is, of course, equal to its energy $\omega^*$).

\subsection{Decay $\rho^0\rightarrow \pi^+\pi^-e^+e^-$}

Our aim here is to derive formulas for the differential branching 
ratios in dilepton momentum and mass $d^2B^{e^+e^-}/dM^2dq^*$ in 
the various formalisms. We will apply the same assumptions and 
approximations which the various authors have used when deriving 
their formulas for the differential cross section. For the
formalisms that satisfy the global variable test 
(\cite{bpp,clgore1}, our formalism), it is sufficient to calculate 
the branching ratio for the virtual photon production
\begin{equation}
\label{virtbrb}
\left(\frac{dB^{\gamma^*}}{dq^*}\right)_M =
\frac{\alpha}{4\pi^2}
\frac{{q^*}^2}{E^*}\int\left(-J^2\right)\ d\Omega_{{\bf q}^{*}}
\end{equation}
[cf. (\ref{virtdecay})] and then multiply it by the function 
${\cal T}(M^2)$, given by Eq.~(\ref{tmsq}). In other cases, the
procedure will be less straightforward. 
 
\begin{figure}
\begin{center}
\leavevmode
\setlength \epsfxsize{8cm}
\epsffile{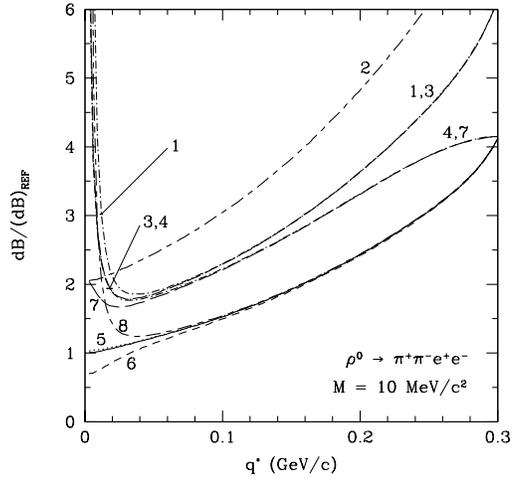}
\end{center}
\caption{The ratio of the differential branching ratio for 
$\rho^0\rightarrow \pi^+\pi^-e^+e^-$ calculated from various 
bremsstrahlung formulas to that of Eq.~(\protect\ref{virtsinger}) 
as a function of the dielectron momentum in the $\rho^0$ rest frame 
at dielectron mass of $M=10$ MeV/$c^2$.
Solid line: formula (\protect\ref{mybremsb});
1: R\"{u}ckl formula; 
2: formula of Goshaw \protect{\it et al.};
3: Gale and Kapusta \protect\cite{galkap} improvement
of R\"{u}ckl formula; 
4: Gale and Kapusta \protect\cite{galkap89} improvement 
of R\"{u}ckl formula; 
5: formula of Balek \protect{\it et al.}, 
identical with \protect$m_T$-scaling supplemented with the real 
photon bremsstrahlung formula; 
6: formula of Cleymans \protect{\it et al.}; 
7: Haglin \protect{\it et al.} improvement of R\"{u}ckl formula;
8: real photon approximation}
\label{comp010}
\end{figure}

\begin{figure}
\begin{center}
\leavevmode
\setlength \epsfxsize{8cm}
\epsffile{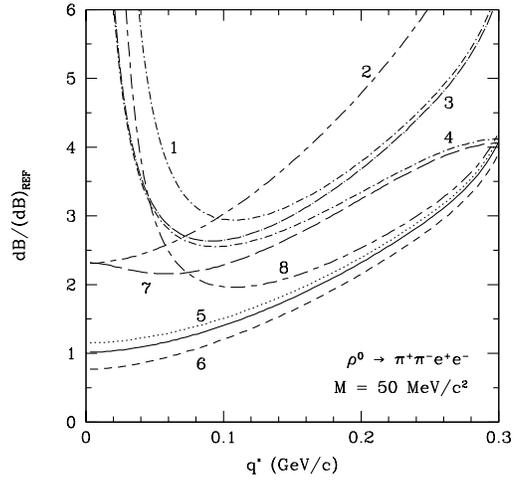}
\end{center}
\caption{Same as Fig.~\protect\ref{comp010} but $M=50$ MeV/$c^2$.}
\label{comp050}
\end{figure}

\begin{figure}
\begin{center}
\leavevmode
\setlength \epsfxsize{8cm}
\epsffile{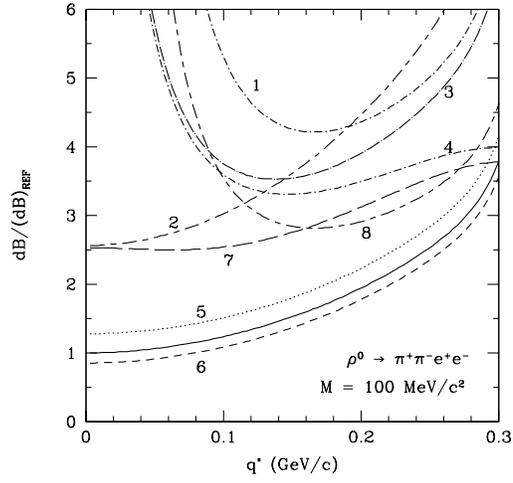}
\end{center}
\caption{Same as Fig.~\protect\ref{comp010} but $M=100$ MeV/$c^2$.}
\label{comp100}
\end{figure}

\begin{figure}
\begin{center}
\leavevmode
\setlength \epsfxsize{8cm}
\epsffile{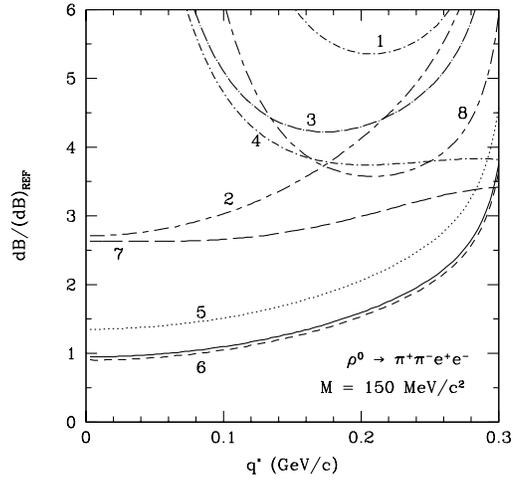}
\end{center}
\caption{Same as Fig.~\protect\ref{comp010} but $M=150$ MeV/$c^2$.}
\label{comp150}
\end{figure}

\begin{figure}
\begin{center}
\leavevmode
\setlength \epsfxsize{8cm}
\epsffile{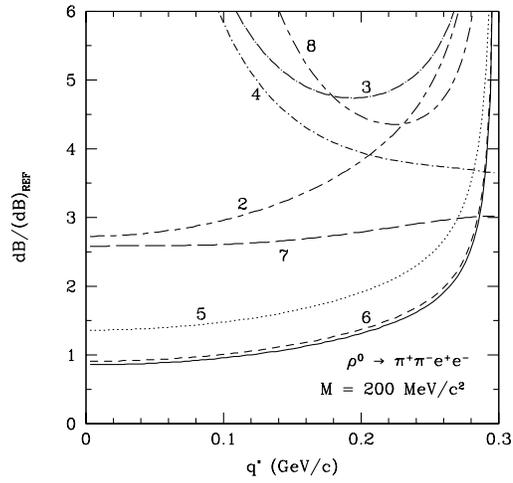}
\end{center}
\caption{Same as Fig.~\protect\ref{comp010} but $M=200$ MeV/$c^2$.
The R\"uckl formula curve is off the scale.}
\label{comp200}
\end{figure}

Applying our formula (\ref{j}) to the process considered here, we 
can write
\begin{equation}
\label{myjvec}
{\bf J}=\frac{2m_\rho^{-1}}{(E^*+M^2/m_\rho)^2-
(v^*q^*\cos\theta^*)^2}
\left[(E^*m_\rho+M^2){\bf v}^{\ *}+
v^*q^*\cos\theta^*\ {\bf q}^{\ *}\right] \ .
\end{equation}
Now we insert this into (\ref{j2}) and calculate the integral which
enters the formula (\ref{virtbrb}). We finish with
\begin{eqnarray}
\label{mybremsb}
\frac{d^2B^{e^+e^-}}{dM^2dq^*}&=&\frac{2\alpha}{\pi}\frac{q^*}{{E^*}}
{\cal T}(M^2)
\left[\frac{1+{v^*}^2-M^2/m_\rho^2}{2v^*(E^*+M^2/m_\rho)}
\ln\frac{E^*+M^2/m_\rho+v^*q^*}{E^*+M^2/m_\rho-v^*q^*}\right. 
\nonumber \\& &-\left.
\frac{(1-{v^*}^2-M^2/m_\rho^2)q^*}{{E^*}^2-(v^*q^*)^2+
(2E^*+M^2/m_\rho)M^2/m_\rho}\right]\ .
\end{eqnarray}
This is the leading term bremsstrahlung formula based on the 
formalism we developed in Sec.~\ref{mesons}. The ratio of 
(\ref{mybremsb}) to the ``exact" formula (\ref{eesinger}) 
is presented as a function of dielectron momentum $q^*$ for 
several dielectron masses in Figs.~\ref{comp010} through 
\ref{comp200} by a solid curve.

The corresponding formula in the Balek, Pi\v{s}\'{u}tov\'{a}, and
Pi\v{s}\'{u}t \cite{bpp} approximation is obtained along the same 
lines. The only difference is in using their (\ref{pisutc}), which, 
translated from the scattering case to the 
$\rho^0\rightarrow\pi^+\pi^-$ decay, reads
\begin{equation}
\label{bppcvec}
{\bf J}=\frac{2E^*}{{E^*}^2-(v^*q^*\cos\theta^*)^2}{\bf v}^{\ *}\ ,
\end{equation}
instead of our (\ref{myjvec}). The result is
\begin{equation}
\label{bppbremsb}
\frac{d^2B^{e^+e^-}}{dM^2dq^*}=\frac{2\alpha}{\pi}\frac{q^*}
{{E^*}^2}{\cal T}(M^2)
\left[\frac{1+{v^*}^2}{2v^*}
\ln\frac{E^*+v^*q^*}{E^*-v^*q^*}-
\frac{\left(1-{v^*}^2\right)E^*q^*}{{E^*}^2-(q^*v^*)^2}\right]\ .
\end{equation}
It is a good check that for $M\ll m_\rho$, Eq.~(\ref{mybremsb}) 
tends to agree with (\ref{bppbremsb}).

When we combined the $M_T$-scaling hypothesis (\ref{mtscaling})
with the leading term bremsstrahlung formula for real photons 
(\ref{myreal}), we got the same result as from the Balek~{\it et~al.}
approximation (\ref{bppbremsb}). 
This prompted us to investigate the connection between 
the two approaches in more detail in Appendix \ref{appmt}.

In order to calculate the differential branching ratio in the 
Cleymans, Goloviznin, and Redlich approximation \cite{clgore1}, 
we also use (\ref{bppcvec}). In addition, we have to discard the dot 
product in Eq.~(\ref{j2}). In this case we get
\begin{equation}
\label{cgrbremsb}
\frac{d^2B^{e^+e^-}}{dM^2dq^*}=\frac{2\alpha}{\pi}\frac{1}{E^*}
{\cal T}(M^2)
\left[\frac{{E^*}^2+(q^*v^*)^2}{2v^*E^*q^*}
\ln\frac{E^*+v^*q^*}{E^*-v^*q^*}-1\right]\ .
\end{equation}

To find the branching ratios in remaining approximations no additional
calculations are needed; we need only to combine the proper formulas.

The R\"{u}ckl approximation formula (\ref{globrueckl}) in terms of
branching ratios may be written as
\begin{equation}
\label{rueglobb}
E^*\frac{d^2B^{e^+e^-}}{dM^2dq^*}= 
\frac{\alpha}{2\pi}\frac{1}{M^2}\sqrt{1-\frac{4\mu^2}{M^2}}\,\, 
q^*\frac{dB^\gamma}{dq^*}.
\end{equation}
Merging this with (\ref{eq7}), we get
\begin{equation}
\label{ruebremsb}
\frac{d^2B^{e^+e^-}}{dM^2dq^*}=
\frac{\alpha^2}{\pi^2}\frac{1}{E^*M^2}
\sqrt{1-\frac{4\mu^2}{M^2}}
\left(\frac{1+{v^*}^2}{2v^*}\ln\frac{1+v^*}{1-v^*}-1\right)\ .
\end{equation}

The Gale and Kapusta \cite{galkap} modification of the 
R\"{u}ckl formula consists in multiplying (\ref{ruebremsb})
by the factor $q^*/E^*$. To account for the later modification by
the same authors \cite{galkap89}, we must also include their 
phase-space correction factor $R(m_\rho^2,M,E^*)$, see (\ref{rgaka}).

To adopt the Haglin, Gale and Emely'anov \cite{hagaem2} modification 
of the R\"{u}ckl formula, we have to multiply (\ref{ruebremsb}) by 
$R(m_\rho^2,M,E^*)(q^*/E^*)^2$.

When dealing with the Goshaw {\it et al.} \cite{goshawee} formalism,
we first rewrite Eq.~(\ref{globgosh}) by means of (\ref{virtbrb}) 
to the form
\begin{equation}
\label{goshb}
\frac{d^2B^{e^+e^-}}{dM^2dq^*}=\frac{2\alpha}{3\pi}\frac{1}{M^2}
\left(1-\frac{\mu^2}{M^2}\right)\sqrt{1-\frac{4\mu^2}{M^2}}
\left(\frac{dB^{\gamma^*}}{dq^*}\right)_M\ .
\end{equation}
The virtual-photon branching ratio on the right-hand side can be 
taken from the Balek {\it et al.} formalism, because they use the
identical four-vector $J$ [compare (\ref{jg}) with (\ref{pisutc})].
It can easily be read off the Eq.~(\ref{bppbremsb}); one simply 
ignores the function ${\cal T}(M^2)$. As a result, we get
\begin{eqnarray}
\label{gosbremsb}
\frac{d^2B^{e^+e^-}}{dM^2dq^*}&=&\frac{4\alpha^2}{3\pi^2}\frac{q^*}
{(E^*M)^2}
\left(1-\frac{\mu^2}{M^2}\right)\sqrt{1-\frac{4\mu^2}{M^2}}
\left[\frac{1+{v^*}^2}{2v^*}
\ln\frac{E^*+v^*q^*}{E^*-v^*q^*}\right.\nonumber \\
&-&\left.\frac{\left(1-{v^*}^2\right)E^*q^*}{{E^*}^2-(q^*v^*)^2}
\right]\ .
\end{eqnarray}

The real photon approximation (\ref{rpa})
combined with (\ref{eesinger}) and (\ref{eq7}) leads to
\begin{equation}
\label{fgabremsb}
\frac{d^2B^{e^+e^-}}{dM^2dq^*}=
\frac{2\alpha}{\pi E^*}{\cal T}(M^2)
\sqrt{1-\frac{4\mu^2}{M^2}}
\left(\frac{1+{v^*}^2}{2v^*}\ln\frac{1+v^*}{1-v^*}-1\right)\ .
\end{equation}

All the formulas we have derived here are normalized to the ``exact"
formula [Eqs.~(\ref{virtsinger}) and (\ref{eesinger})] and 
visualized as functions of dielectron momentum at fixed dielectron 
masses in Figs.~\ref{comp010} through \ref{comp200}. 

\section{COMMENTS AND CONCLUSIONS}
\label{comments}

We based our formalism for dilepton production via virtual photon
bremsstrahlung on the following assumptions:
\begin{enumerate}
\item
Only the Feynman diagrams in which a virtual photon line is attached 
to one of the external legs are important.
\item
The charged particles that participate in the process are
pseudoscalar mesons or unpolarized spin-one-half fermions.
\item
The dilepton four-momentum in the argument of the Dirac 
$\delta$-function in Eq.~(\ref{xsec}) (or in similar relations for 
decays or processes with more than two particles in the initial 
state) may be neglected.
\item
The modifications of electromagnetic interactions of hadrons induced 
by form factors and anomalous magnetic moments (in the case of 
fermions) are negligible.
\end{enumerate}

To obtain the leading term approximation, we further neglected terms 
that are proportional to the derivatives of the nonradiative matrix 
element squared. Our notion of the leading term approximation is thus 
based on the proportionality of the bremsstrahlung cross section to 
that of the nonradiative reaction rather than on the order in dilepton 
momentum $q$. In our leading term formulas, we use the four-vector $J$ 
that contains also subleading terms in $q$. The numerical comparison 
with the ``exact" formula shows that it is beneficial. 

In Sec.~\ref{survey}, we have pointed out the additional assumptions 
which are required to reduce our formulas to the virtual 
bremsstrahlung formulas that were derived or suggested previously. Of
these formulas, only one, namely that of Balek, Pi\v{s}\'{u}tov\'{a} 
and Pi\v{s}\'{u}t \cite{bpp}, satisfied both the Lorentz covariance 
and global variable tests.

The numerical analysis presented in Sec.~\ref{bremsstrahlung} shows
that all leading term virtual bremsstrahlung formulas (including 
ours) overestimate the dilepton production, albeit to different 
degrees. As a matter of fact, also the real photon 
production is overestimated by the standard leading-term formula 
(see Figs.~\ref{realgam} and \ref{comp000}). 
But due to the more complex nature of virtual bremsstrahlung, 
the situation with dileptons is more complicated.
Besides the excess in the high-momentum region, which is similar to 
that for real photons, some formulas also overshoot the yield at small
momenta. These formulas relate dilepton production to
the real photon cross section rather than to the virtual one. Near 
threshold, the differential cross section in virtual photon momentum 
behaves like ${q^*}^2$ [in both the ``exact" formula 
(\ref{virtsinger}) and Eq.~(\ref{mybremsb})], whereas for real photons
the behavior is ${q^*}^{-1}$ [see Eqs.~(\ref{realsinger}) and 
(\ref{eq7})].

Energy-momentum conservation is not built into the leading term 
bremsstrahlung formulas. The fact that they overestimate the dilepton
yield at high dilepton momenta is therefore not surprising. But we 
will see that it is only a part of the story.

\begin{figure}
\begin{center}
\leavevmode
\setlength \epsfxsize{8cm}
\epsffile{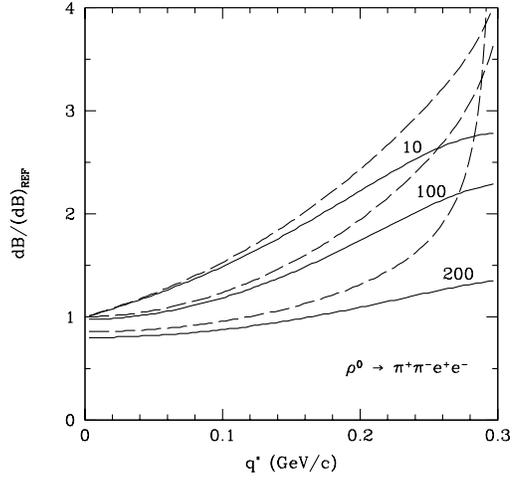}
\end{center}
\caption{The role of the phase-space correction factor 
(\protect\ref{rgaka}) in virtual bremsstrahlung at three different
dilepton masses $M$: 10 MeV/$c^2$ (upper), 100 MeV/$c^2$
(medium), and  200 MeV/$c^2$ (lower). The leading term
bremsstrahlung formula (\protect\ref{mybremsb}) with (solid) and
without (dashed) the correction factor was divided by 
(\protect\ref{eesinger}).}
\label{myform}
\end{figure}

Gale and Kapusta \cite{galkap89} introduced the factor (\ref{rgaka}), 
which accounts for the shrinking of the final-state phase space and 
prevents violation of energy-momentum conservation. They, and also 
the authors of 
a later work \cite{hagaem2}, combined it with the R\"{u}ckl formula. 
This procedure obscures the conclusions somewhat, because the 
R\"{u}ckl formula contains a wrong numerical
factor. We therefore apply the correction factor (\ref{rgaka}) to
the leading term virtual bremsstrahlung formula (\ref{mybremsb}). 
The results are displayed in Fig.~\ref{myform} after being normalized 
to the ``exact" formula. Comparison with corresponding non-corrected 
curves shows that only a part of the excess over the ``exact" formula 
has been removed. The remaining excess should be ascribed to other 
sources. It can be a modification of the nonradiative matrix element 
combined with a destructive interference between radiation from the 
external particles and from the contact term. Both these effects are 
neglected in the leading-term approximation.

The bremsstrahlung calculations play an important role in assessing 
the conventional sources of photons or dileptons in experiments 
aimed at revealing anomalous production of electromagnetic probes as 
a sign of new physics phenomena. For example, Haglin and Gale
\cite{haga94} have recently shown that bremsstrahlung is the 
largest source of low-mass dielectrons in 4.9 GeV $pp$ collisions. 
A quite good, even if not perfect, agreement with experimental data 
\cite{dls} was achieved. The authors used the modified R\"{u}ckl 
formula, discussed in \ref{modifications}. But our toy example showed 
that this formula overestimates the dielectron yield by a factor of 
2--3. This suggests that calculations using a more correct 
bremsstrahlung formalism have to be performed before a definite 
conclusion about the physics behind the Dilepton
Spectrometer collaboration data \cite{dls} is drawn.

To conclude, we have derived formulas for differential cross 
sections of dilepton production via virtual photon bremsstrahlung 
from pseudoscalar mesons and unpolarized fermions in the leading and
subleading approximations. These formulas satisfy the conditions 
imposed by Lorentz covariance and gauge invariance.
From the practical point of view, the leading term formulas 
(\ref{d6sig}), (\ref{d6sigmix}), and (\ref{newglobal}) are the most 
important. They enable us to estimate the radiative cross sections on 
the basis of the corresponding nonradiative cross section. Comparing 
to previously published formulas, our formalism exhibits the best 
agreement with
the exact formula when applied to a concrete physical process.

\acknowledgements

The author is indebted to V.~Balek, J.~Cleymans, C.~Gale, 
K.~Haglin, J.~Kapusta, J.~Pi\v{s}\'{u}t, M.~Prakash, and 
K.~Redlich for stimulating discussions and critical remarks. 
M.~Prakash also read the manuscript very carefully and suggested 
many improvements of the text. My interest in virtual
bremsstrahlung was aroused in discussions with J.~Anto\v{s}, 
C.~Fabjan, J.~Schukraft, and J.~Thompson, the members of the 
HELIOS collaboration at CERN. H.~Eggers corrected several typos.
The stay at the State University of New York at Stony Brook 
was supported by the U.S. Department of Energy under grant 
No.~DE-FG02-88ER-40388. A part of the work was done during 
a visit to the CERN Theory Division, the 
hospitality of which is gratefully acknowledged.

\appendix

\section{\protect{$M_T$}-scaling and 
virtual bremsstrahlung formalism}
\label{appmt}

Let us describe the multiparticle production in a frame where the 
velocities of the two incident particles are collinear and define 
a collision axis. The dot product between the dilepton momentum 
${\bf q}\ $ and the momentum 
${\bf p}_i\ $ of the $i$th particle can then be written as
\begin{equation}
\label{qdotp}
{\bf q}\!\cdot\!{\bf p}_i\ = |{\bf q}|p_{i,L}\cos\theta
+{\bf q}_T\!\cdot\!{\bf p}_{i,T}\ ,
\end{equation}
where $\theta$ is the angle between the dilepton momentum and the 
projectile velocity.

If the transverse momenta of the outgoing charged particles are
negligibly small (the criterion of negligibility depends, of course,
on the value of $\theta$), the rightmost term in (\ref{qdotp}) can be
omitted. As a consequence, the virtual photon bremsstrahlung formula 
in the Balek {\it et~al.} approximation (\ref{bppn}) can be rearranged
into 
\begin{equation}
\label{a1}
E\left(\frac{d^3\sigma^{\gamma^*}}{d^3q}\right)_M
=\frac{\alpha}{4\pi^2E^2}\int\ -\left(\sum \frac{Q_i^\prime}{E_i}
\ \frac{p_i^\mu}
{1-E^{-1}v_{i,L}|{\bf q}|\cos\theta\ } \right)^2d\sigma_0\ 
\end{equation}
with $v_{i,L}=p_{i,L}/E_i$. Let us introduce a new vector 
${\bf p}\ $ by
\begin{eqnarray}
\label{a2}
p_L&=&q_L\nonumber\\
{\bf p}_T&=& \frac{{\bf q}_T}{q_T}\ \sqrt{q_T^2+M^2}
\end{eqnarray}
and denote its polar angle by $\alpha$. We have
\begin{eqnarray}
\label{a3}
{\bf p}^{\ 2}=q_L^2+q_T^2+M^2=E^2\ ,\nonumber\\
\cos\alpha=\frac{p_L}{|{\bf p}|}=\frac{|{\bf q}|}{E}\cos\theta\ .
\end{eqnarray}
Equation (\ref{a1}) then reads
\begin{equation}
\label{a4}
E\left(\frac{d^3\sigma^{\gamma^*}}{d^3q}\right)_M
=\frac{\alpha}{4\pi^2{\bf p}^2}\int\ -\left(\sum \frac{Q_i^\prime}
{E_i}\ \frac{p_i^\mu}
{1-v_{i,L}\cos\alpha\ } \right)^2d\sigma_0\ .
\end{equation}
The right-hand side is nothing else but the invariant cross section
(\ref{myreal}) for producing a real photon with the momentum 
${\bf p}\ $ and energy $\omega=|{\bf p}|$ in the same approximation 
(negligible transverse momenta of hadrons). We have thus recovered 
the $M_T$-scaling hypothesis (\ref{mtscaling}).

The $M_T$-scaling is clearly an approximate phenomenon. To arrive at 
it, we first made the approximations recapitulated in 
Sec.~\ref{comments}, 
that led to our basic formulas of Sec.~\ref{mesons}. Then we 
added another approximation
to get the Balek {\it et al.} formula, which is for negligible
transverse momenta equivalent to merging
$M_T$-scaling with leading term formula for bremsstrahlung of real
photons. In realistic situations, however, the violation 
coming from nonvanishing transverse momenta of hadrons also
plays an important role. In our simple example--the nonradiative
decay $\rho^0\rightarrow\pi^+\pi^-$ discussed in 
Sec.~\ref{bremsstrahlung}, the outgoing hadrons do not have 
any transverse momenta and the two approaches are fully equivalent.

\section{``Exact" formula for the 
$\rho^0\rightarrow\pi^+\pi^-\gamma^*$ branching ratio}
\label{appsingvirt}
The matrix element is given by the Feynman diagrams depicted
in Fig.~\ref{feynman} with a real photon replaced by a massive one. 
We define the invariant variables
\begin{eqnarray}
\label{stu}
s &=& (p_1+p_2)^2\ ,\nonumber\\
t^\prime &=& (p_2+q)^2-m_\pi^2\ ,\nonumber\\
u^\prime &=& (p_1+q)^2-m_\pi^2\ ,
\end{eqnarray}
with $p_1$, $p_2$, and $q$ being the four-momenta of $\pi^+$, $\pi^-$,
and $\gamma^*$, respectively. The invariants (\ref{stu}) satisfy the
usual relation
\begin{equation}
\label{stucond}
s+t^\prime+u^\prime=m_\rho^2+M^2\ .
\end{equation}
The sum over the $\rho^0$ and $\gamma^*$ polarizations of the matrix 
element squared is equal to
\begin{equation}
\label{virtmesq}
\sum_{\lambda_\rho,\lambda_{\gamma^*}}\left|{\cal M}\right|^2=
\left(\frac{ef_{\rho\pi\pi}}{t^\prime u^\prime}\right)^2\ 
A^{\mu\alpha}A^{\nu\beta}
\sum_{\lambda_\rho}\epsilon_\alpha(\lambda_\rho)
\epsilon_\beta^*(\lambda_\rho)
\sum_{\lambda_{\gamma^*}}\epsilon_\mu(\lambda_{\gamma^*})
\epsilon_\nu^*(\lambda_{\gamma^*})\ ,
\end{equation}
with the tensor $A$ given by
\begin{equation}
\label{amualpha}
A^{\mu\alpha}=t^\prime(2p_1+q)^\mu(p_1-p_2+q)^\alpha+
u^\prime(2p_2+q)^\mu(p_2-p_1+q)^\alpha-2t^\prime u^\prime 
g^{\mu\alpha}\ .
\end{equation}
Thanks to the relations $q_\mu A^{\mu\alpha}=0$ and 
$A^{\mu\alpha}P_\alpha=0$, where $P$ is the $\rho^0$ four-momentum, 
the sums of products of polarizations vectors in (\ref{virtmesq}) 
can be replaced by the corresponding metric tensors. After a little 
algebra, we get
\begin{equation}
\sum_{\lambda_\rho,\lambda_{\gamma^*}}\left|{\cal M}\right|^2=
16\pi\alpha f_{\rho\pi\pi}^2\left[2+
a_1\left(\frac{1}{t^\prime}+\frac{1}{u^\prime}\right)
-a_2\left(\frac{1}{{t^\prime}^{\ 2}}+\frac{1}{{u^\prime}^{\ 2}}
\right)\right]\ ,
\end{equation}
where
\begin{eqnarray}
\label{dcka}
a_1 &=& \frac{1}{E^*}\left[E_m^*\left(m_\rho^2-2m_\pi^2-2m_\rho E^*
\right)+M^2\left(E_m^*+m_\rho/4\right)\right], \nonumber\\
a_2 &=& \frac{1}{4}\left(m_\rho^2-4m_\pi^2\right)
\left(4m_\pi^2-M^2\right).
\end{eqnarray}
See also (\ref{estm}). To calculate the invariant decay width of the 
unpolarized $\rho^0$, namely 
\begin{equation}
\label{invdecw}
E\left(\frac{d^3\Gamma_{\rho^0\rightarrow\pi^+\pi^-\gamma^*}}{d^3q}
\right)_M=\frac{\pi}{6m_\rho}\int
\delta(P-q-p_1-p_2)
\sum_{\lambda_\rho,\lambda_{\gamma^*}}\left|{\cal M}\right|^2
\prod_{i=1}^2\frac{d^3p_i}{2E_i(2\pi)^3}
\end{equation}
we need integrals of the type
\begin{equation}
\label{in}
I_n=\int\frac{d^3p_1}{E_1}\int\frac{d^3p_2}{E_2}\delta(P-q-p_1-p_2)
\frac{1}{{t^\prime}^{\ n}}\ ,
\end{equation}
which can be computed most simply in the two-pion rest frame after
the substitutions $Q=p_1+p_2$ and $R=p_1-p_2$. We get
\begin{eqnarray}
\label{i012}
I_0 &=& 2\pi\xi\ ,\nonumber\\
I_1 &=& \frac{\pi}{m_\rho q^*}\ln\frac{E^*+\xi q^*}{E^*-\xi q^*}\ ,
\nonumber\\
I_2 &=& \frac{2\pi\xi}{m_\rho^2}\frac{1}{{E^*}^2-(\xi q^*)^2}\ ,
\end{eqnarray}
with $\xi$ given by Eq.~(\ref{ximsq}). Putting it all together and
using the formula \cite{singer}
\begin{equation}
\label{rhodecw}
\Gamma_{\rho\rightarrow\pi\pi}=\frac{f_{\rho\pi\pi}^2}
{48\pi m_\rho^2}\left(m_\rho^2-4m_\pi^2\right)^{3/2}\ ,
\end{equation}
we  eventually get (\ref{virtsinger}).

\section{Leading term bremsstrahlung and event generators}
\label{appleading}
The attitude to the bremsstrahlung in hadronic reactions has been 
much influenced by the fact that the leading term in quantum 
electrodynamics is equivalent to the corresponding classical 
expression \cite{jackson}.
Similar relation exists also for virtual bremsstrahlung \cite{bpp}.

As a consequence, also the event generators constructed so far were, 
according to our knowledge, ``classical." The configuration of 
hadrons in momentum space was considered a source of photons or 
dileptons. The momenta of hadrons were assumed untouched by the 
creation of a photon or dilepton. 
This led to problems with the energy-momentum conservation.

We think that those problems can be cured by a more quantal approach. 
One should  consider the transition probability from the initial 
state to the ``complete" final state, containing both hadrons and 
a photon (or a dilepton). The energy-momentum conservation is firmly 
enforced. As we show below, the leading term bremsstrahlung 
approximation means that the probability
of such a transition can be expressed as a product of two terms. The
first of them is assumed to be independent of photon or dilepton
four-momentum, the second one contains the momenta of both hadrons and
a photon (a dilepton), but in a simple way. What is approximate is
the transition probability (rephrased into the cross section, decay
width, or other observable quantities), not the energy-momentum 
conservation.

To illustrate our point in more detail, let us return back to the 
basic equations. To simplify the discussion, we will consider the 
reaction
\begin{equation}
\label{photonprod}
a+b\rightarrow 1+2+\cdots+n+\gamma\ .
\end{equation} 
Its cross section is given by (we choose the center-of-mass reference
frame)
\begin{equation}
\label{photonxsec}
\sigma = \frac{1}{4p_a^*\sqrt{s}}
\int\sum_{\lambda_\gamma}\left|{\cal M}\right|^2
(2\pi)^4\delta(p_a+p_b-\sum_i p_i-q)
\frac{d^3q}{2\omega(2\pi)^3}\ 
\prod_{i=1}^n\frac{d^3p_i}{2E_i(2\pi)^3}\ .
\end{equation}
Due to the presence of the four-dimensional $\delta$-function, only 
the $(3n\!-\!1)$ momentum components are independent. It is 
convenient to transform the integration region in (\ref{photonxsec}) 
into a $(3n\!-\!1)$-dimensional unit cube. Several such procedures 
exist in the literature (see, e.g., 
\cite{vanhove,pene,kittel,jadach}). We thus get
\begin{equation}
\label{photonxsecx}
\sigma =\int\frac{d^{3n-1}\sigma}{d^{3n-1}\xi}\ d^{3n-1}\xi
\end{equation}
with
\begin{equation}
\label{photonxsecxi}
\frac{d^{3n-1}\sigma}{d^{3n-1}\xi} = \frac{1}{4p_a^*\sqrt{s}}
\sum_{\lambda_\gamma}\left|{\cal M}\right|^2
f(\xi_1,\xi_2,\cdots,\xi_{3n-1})\ .
\end{equation}
The function $f$ results from the integration over the four dependent
variables and from the Jacobian of the substitutions
\begin{eqnarray}
\label{transfp}
{\bf p}_i &=& {\bf p}_i(\xi_1,\xi_2,\cdots,\xi_{3n-1}),\hspace{1in} 
i=1,2,\cdots,n \ , \\
\label{transfq}
{\bf q} &=& {\bf q}(\xi_1,\xi_2,\cdots,\xi_{3n-1})\ .
\end{eqnarray}
The momenta given by Eqs.~(\ref{transfp}) and (\ref{transfq}) satisfy 
the energy-momentum conservation in reaction (\ref{photonprod}). The 
above substitutions are not straightforward, they usually require to 
solve an algebraic equation. We refer the reader to the original 
literature \cite{vanhove,pene,kittel,jadach}.

The master equation of the event generator is equivalent to the 
evaluating of the cross section by a Monte Carlo method:
\begin{equation}
\label{mci}
\sigma=\frac{1}{N}\sum_{k=1}^{N}
\left(\frac{d^{3n-1}\sigma}{d^{3n-1}\xi}\right)_{\vec \xi^{(k)}}\ ,
\end{equation}
where the sums runs over the random points uniformly distributed
within the $(3n\!-\!1)$-dimensional unit cube. Each such point
generates an ``event"--a set of momenta of particles in the final
state. The quantity (\ref{photonxsecxi}) is a weight that is assigned
to each ``event". If we succeeded in finding such a form of
substitutions (\ref{transfp}) and (\ref{transfq}) that the weights
of all events are same, we would have an ideal event generator.

We have not made any approximation yet, Eqs.~(\ref{photonxsecx}) and 
(\ref{photonxsecxi}) 
are exact. Now, we are again going to 
consider only the radiation coming from the external particles and to
ignore the possible changes in the strong ``core" of the Feynman  
diagram if one of the external legs goes off-shell. The sum of the 
matrix 
element squared over the photon polarizations is then simply equal to
\begin{equation}
\label{mephot}
\sum_{\lambda_\gamma}\left|{\cal M}\right|^2=4\pi\alpha
\left|{\cal M}_0\right|^2\left(-J_R^2\right)\ ,
\end{equation}
where $J_R$ is given by Eq.~(\ref{jr}).
To relate $|{\cal M}_0|^2$ to observable quantities, we need a set of 
hadron momenta that satisfy the energy-momentum conservation for the 
nonradiative reaction (\ref{reaction1}). In other words, for the sake
of evaluation of the integrand in (\ref{photonxsecxi}) we must 
``spoil" the momenta ${\bf p}_i$ a little. This procedure replaces the 
assumption that the four-vector $q$ in the argument of the 
$\delta$-function in (\ref{xsec}) may be neglected. We can change the 
hadron momenta in many different ways. Here is one of them. 

We first express $\xi_{3n-3}$,
$\xi_{3n-2}$, and $\xi_{3n-1}$ in terms of ${\bf q}$ and remaining
$\xi$'s by inverting Eq.~(\ref{transfq}). The momenta of hadrons thus
become functions of 
$\xi_1$, $\xi_2$, $\cdots$, $\xi_{3n-4}$, and ${\bf q}$.
Now we can define a set of hadron momenta
\begin{equation}
\label{pzero}
{\bf p}^{(0)}_i=\lim_{{\bf q}\rightarrow {\bf 0}}
{\bf p_i}(\xi_1,\xi_2,\cdots,\xi_{3n-4},{\bf q})\ ,
\end{equation}
which satisfy the energy-momentum conservation for reaction 
(\ref{reaction1}). We can use them to evaluate (\ref{mephot}) 
by means of the purely hadronic matrix element, which is related
to the cross section of reaction (\ref{reaction1}). The weight of the
$k$th event thus becomes
\begin{equation}
\label{wk}
w_k=\frac{4\pi\alpha}{4p_a^*\sqrt{s}}\left|{\cal M}_0
\right|^2_{\left\{{\bf p}_i^{(0)}\right\}_k}\left(-J_R^2\right)_
{\left\{{\bf p}_i,
{\bf q}\right\}_k}f\left(\xi_1^{(k)},\dots,\xi_{3n-1}^{(k)}\right)\ .
\end{equation} 
The above procedure can be used if the exclusive cross section is 
given by an analytic formula. 

In a more realistic situation, the bremsstrahlung cross section is 
estimated from the experimental data on the nonradiative reaction 
(\ref{reaction1}) on event-by-event basis. The Nature acts as 
an ``ideal event generator", which generates exclusive sets of
hadron momenta ${\bf p}_i^{(0)}$, $i=1,\cdots,n$. Each set is
assigned the same weight $\sigma_0$. The distribution of events in
hadron momenta is governed by the matrix element squared 
$\left|{\cal M}_0\right|^2$ and the phase space. The role of the
latter can again be described using the substitutions
\begin{equation}
\label{transfp0}
{\bf p}_i^{(0)} = {\bf p}_i^{(0)}(\xi_1,\xi_2,\cdots,\xi_{3n-4})\ , 
\hspace{1in}i=1,2,\cdots,n \ 
\end{equation}
that guarantee the energy-momentum conservation in reaction 
(\ref{reaction1}).
For properly chosen substitutions (\ref{transfp0}), the 
$\xi$-dependence of the partly integrated Jacobian $f^{(0)}$ 
compensates the ${\bf p}$-dependence of the 
matrix element squared in the sense that the product of them is 
constant. The weight of each event thus is, as required,
\begin{equation}
\label{nature}
\sigma_0=\frac{1}{4p_a^*\sqrt{s}}\left|{\cal M}_0\right|^2
f^{(0)}\left(\xi_1,\cdots,\xi_{3n-4}\right)\ .
\end{equation}
The momentum distribution of experimental events corresponds to 
a uniform distribution of points within
the $(3n\!-\!4)$-dimensional unit cube. Their coordinates are given 
by the inverse substitution
\begin{equation}
\label{xicoo}
\xi_i=\xi_i\left({\bf p}_1^{(0)},
{\bf p}_2^{(0)},\cdots,{\bf p}_n^{(0)}
\right)\ ,\hspace{1in}i=1,\cdots,3n-4\ .
\end{equation}

To generate a photon, we proceed in two steps. Firstly,
for each event, we have to ``spoil" the hadron momenta 
and add a momentum of photon in such a way that they together cope 
with the energy-momentum constraints for the 
reaction~(\ref{photonprod}). 
Secondly, we have to find the weight of this photonic event.

In order to obtain the momenta in the 
radiative event, we supplement the coordinates (\ref{xicoo}) by 
three random numbers 
$\xi_{3n-3}$, $\xi_{3n-2}$, and $\xi_{3n-1}$. The ``spoilt" hadron
momenta and the photon momentum are now given
by Eqs.~(\ref{transfp}) and (\ref{transfq}). 
The weight of an event can be found by inserting 
$\left|{\cal M}_0\right|^2$ from Eq.~(\ref{nature}) to 
Eq.~(\ref{wk}).
\begin{equation}
\label{wkprime}
w_k^\prime=4\pi\alpha\ \sigma_0
\left(-J_R^2\right)_{\{{\bf p}_i,{\bf q}\}_k}\frac
{f\left(\xi_1^{(k)},\dots,\xi_{3n-1}^{(k)}\right)}
{f^{(0)}\left(\xi_1^{(k)},\dots,\xi_{3n-4}^{(k)}\right)}\ .
\end{equation}
Of course, the substitutions (\ref{transfp}) and (\ref{transfp0}) 
cannot be independent. They must satisfy the condition (\ref{pzero}).
In actual calculation, we do not usually know the ideal substitution 
(\ref{transfp0}) for momenta in the nonradiative reaction,
and work with a substitution that gives fluctuating right-hand
side of Eq.~(\ref{nature}). The weights of the photon-producing events
(\ref{wkprime}) should be influenced only little, because they contain
the ratio $f/f^{(0)}$. 

The dilepton generators in either  global variables or momenta of
leptons can be constructed along the same lines with obvious 
modifications given by the different number of variables and 
different cross sections.

\begin{references}
\bibitem{invint}
The derivation can also be found, e.g., in \cite{bpp}.
\end{references}
\end{document}